\documentclass[final,5p]{elsarticle}
\pdfoutput=1 
\usepackage[switch]{lineno}
\usepackage[x11names]{xcolor}
\usepackage[colorlinks,linkcolor=blue,allcolors=blue]{hyperref}
\AtBeginDocument{\hypersetup{citecolor=DodgerBlue4}}
\usepackage{amsmath}
\usepackage{amssymb}
\usepackage{gensymb}
\usepackage{amstext}
\usepackage{mathtools}
\usepackage{esint}
\usepackage{caption}
\usepackage{wrapfig}
\usepackage{float}
\usepackage{dblfloatfix}
\usepackage[symbol]{footmisc}
\journal{Nuclear Instrumentation and Methods}
\usepackage{geometry}
\usepackage{upgreek}
\usepackage{epsfig}
\usepackage{subcaption}

\usepackage{array}
\usepackage{xcolor}
\usepackage{multirow} 
\usepackage{rotating}
\usepackage{amsfonts}
\usepackage{booktabs}
\usepackage{siunitx}
\usepackage{chemformula}
\newcolumntype{P}[1]{>{\centering\arraybackslash}p{#1}}
\usepackage{amsmath}
\usepackage{makecell}
\usepackage{titlesec}

\titleformat{\chapter}
  {\normalfont\fontsize{16}{19}\bfseries}
  {\thechapter}
  {1em}
  {}

\titleformat{\section}
  {\normalfont\fontsize{12}{17}\bfseries}
  {\thesection}
  {1em}
  {}

\titleformat{\subsection}
  {\normalfont\fontsize{11}{16}\bfseries\slshape}
  {\thesubsection}
  {1em}
  {}

\titleformat{\subsubsection}
  {\normalfont\fontsize{10}{12}\bfseries\slshape}
  {\thesubsubsection}
  {1em}
  {}

\begin{document}


\begin{frontmatter}

\title{Modeling Athermal Phonons in Novel Materials using the G4CMP Simulation Toolkit}


\author[1,2]{I. Hernandez\corref{cor1}}
\author[2]{R. Linehan\corref{cor1}}
\author[1,2]{R. Khatiwada}
\author[1,2]{K. Anyang}
\author[2,4]{D. Baxter}
\author[2,4]{G. Bratrud}
\author[4,2]{E. Figueroa-Feliciano}
\author[2]{L.  Hsu}
\author[3]{M. Kelsey}
\author[2]{D. Temples}
\cortext[cor1]{Corresponding authors: ihernandez6@hawk.iit.edu and linehan3@fnal.gov}

\address[1]{Illinois Institute of Technology, Department of Physics, Chicago, IL 60616, USA}
\address[2]{Fermi National Accelerator Laboratory Laboratory, Batavia, IL 60510, USA}
\address[3]{Texas A\&M University, 400 Bizzell St. College Station, TX 77843}
\address[4]{Department of Physics \(\&\) Astronomy, Northwestern University, Evanston, IL 60208, USA}



\begin{abstract}

Understanding phonon and charge propagation in superconducting devices plays an important role in both performing low-threshold dark matter searches and limiting correlated errors in superconducting qubits. The Geant4 Condensed Matter Physics (G4CMP) package, originally developed for the Cryogenic Dark Matter Search (CDMS) experiment, models charge and phonon transport within silicon and germanium detectors and has been validated by experimental measurements of phonon caustics, mean charge-carrier drift velocities, and heat pulse propagation times. In this work, we present a concise framework for expanding the capabilities for phonon transport to a number of other novel substrate materials of interest to the dark matter and quantum computing communities, including sapphire (Al\(_{2}\)O\(_{3}\)), gallium arsenide (GaAs), lithium fluoride (LiF), calcium tungstate (CaWO$_{4}$), and calcium fluoride (CaF\(_{2}\)). We demonstrate the use of this framework in generating phonon transport properties of these materials and compare these properties with experimentally-determined values where available.
\end{abstract}

\begin{keyword}
Phonon \sep Simulation \sep G4CMP \sep Dark Matter \sep Material Science \sep Qubits \sep Sapphire
\end{keyword}

\end{frontmatter}


\sloppy
\section{Introduction}
\label{sec:intro}

Mounting astrophysical and cosmological observations indicate the existence and abundance of a cold, frictionless nonbaryonic ``dark'' matter in the universe \cite{Arbey,Cold_Dark_Matter,Cold_DM_2}. As direct detection searches for particle-like candidates with masses above the GeV/c\(^{2}\) scale have yielded only null results in the past few decades~\cite{LZ_Experiment,ADMX_Experiment}, there has recently been increased interest in developing low-energy threshold detectors capable of probing parameter space at lower masses, down to the keV/c\(^{2}\) scale~\cite{SnowmassLowThresholdReport}. A common class of such detectors rely on cryogenic (mK) solid-state substrates coupled to superconducting films, in which energy depositions in the substrate generate phonon excitations that can be sensed in the films \cite{TES_DETECTOR,KIPM_Dylan,linehan2024estimating,SQUAT}. These phonons may be generated from a number of different potential interaction types: direct nuclear recoils of dark matter on lattice ions, phonons produced from electron-hole recombination after a dark matter ionization event, and Neganov-Trofimov Luke phonons from drifting freed electrons through the crystal using an electrostatic field. The phonon response of silicon and germanium under these effects has been well-studied \cite{Neganov_1,Neganov_2,NRIonizationYieldSCDMS,IonizationYieldGermanium}.


In recent years there has been a broadening of the landscape of materials proposed for phonon-mediated cryogenic dark matter search experiments. While historically these detectors have been made from silicon, germanium, calcium tungstate (CaWO\(_{4}\)), or calcium fluoride (CaF\(_{2}\))~\cite{Agnese_2019,SuperCDMs,CRESSTIII,Ishidoshiro}, future dark matter detectors may benefit from other carefully chosen novel target materials. For example, polar crystals like sapphire (Al\(_{2}\)O\(_{3}\)) and gallium arsenide (GaAs) increase sensitivity to DM candidates coupling to the dark photon. These materials, especially those with many accessible optical modes, give sufficiently strong directional dependence of DM scattering rates to enable daily modulation searches, increasing DM discovery power~\cite{Griffin_2018,Knapen_2018}. A thorough grasp of (and ability to model) phonon propagation in these materials will be required for signal- and background-model building in these experiments.

Moreover, some of these materials (like sapphire) are commonly used in the quantum computing community as chip substrates due to their low dielectric losses~\cite{Sapphire_Quantum_Computing}. As superconducting qubits have been shown to experience catastrophic correlated errors due to the presence of phonons produced by ionizing radiation, quantitatively understanding phonon transport in such chips may facilitate developing strategies that mitigate such errors \cite{Martinis,Arno,mcewen2024,Downconversion}.


Modeling phonon propagation in novel materials is therefore a foundation for studying signals and backgrounds in particle-like dark matter searches and sources of correlated errors in superconducting qubits. The Geant4 Condensed Matter Physics (G4CMP) package is an open-source particle-tracking simulation tool built to perform this modeling, originally in the context of low-mass dark matter experiments \cite{Kelsey_2023}. While it successfully models phonon transport within silicon and germanium substrates, it currently does not include information for phonon propagation in other target materials. As other substrate materials are adopted in dark matter searches or quantum computing, a thorough understanding of chip response to energy depositions will rely on a corresponding expansion of the set of materials in which G4CMP can model phonon transport. 

In this work we achieve an expansion of G4CMP's phonon-modeling capability for a large number of materials of interest to the low-mass dark matter and quantum information communities, and provide a framework for extending this to further materials. In Section~\ref{sec:Phonon_Kinematics}, we discuss the phonon physics that underlies G4CMP's transport modeling and present the set of parameters needed for modeling phonon response in a new material. In Section~\ref{sec:ModelResultsAndValidation}, we calculate this set of parameters for several new materials, demonstrate successful integration into G4CMP, and validate the results against experimental results available in the literature. We discuss potential extensions to this modeling and its impact on dark matter detection and quantum computing efforts in Section~\ref{sec:Discussion}, and conclude in Section~\ref{sec:Conclusions}.

\section{Phonon Kinematics}
\label{sec:Phonon_Kinematics}

When high-energy phonons are created from energy deposits in a chip, the system evolves in a way governed by several distinct phonon kinematics processes. We consider as an example an optical phonon created from a DM scatter in the substrate. First, this phonon can undergo anharmonic decay into multiple lower-energy phonons via one of several possible channels, including the Klemens channel\footnote{Here, LO indicates longitudinal optical phonon modes, TO indicates transverse optical modes, and XA indicates either longitudinal or transverse acoustic modes. In the acoustic mode, we keep this general as arbitrary combinations may be energetically prohibited in certain materials depending on the phonon dispersion.} (LO\(\rightarrow\)XA+XA), the Ridley Channel (LO\(\rightarrow\)TO+XA), and the Valée-Bogani channel (LO\(\rightarrow\)LO+XA)~\cite{PhysRevB.69.235208}. All of these optical phonon decay channels occur on $\mathcal{O}(ps)$ timescales. Once downconversion from the optical branch phonons to acoustic branch phonons occurs, there is further downconversion of the acoustic phonons, fragmenting the initial phonon energy even more. Scattering on isotopic impurities may change acoustic phonon's trajectories and polarizations, and has been noted to possibly also play a role in phonon thermalization~\cite{AthermalPhononDownconversionTheory}. Selection of the proper final-state phonon polarizations during downconversion and scattering is governed by the fractional density of states (DOS) for the various possible phonon polarizations. At sufficiently low acoustic phonon energies, the scattering and downconversion mean free paths may extend beyond the substrate dimensions, and the phonon trajectory becomes limited by boundary scatters (i.e. enters the ``ballistic regime''). At all energies, phonon propagation is governed by the substrate's crystal structure, leading to formation of caustics,  preferred directions into which phonons' group velocities are focused by the crystal anisotropy. G4CMP simulates the subset of these physical processes involving acoustic phonons: acoustic phonon anharmonic downconversion, isotopic scattering, and transport along caustics.\footnote{While scattering and downconversion of phonons at surfaces is simulated in G4CMP, the complexity of that topic puts it beyond the scope of this work. Optical phonons are not simulated in G4CMP due to a combination of factors: not only are these phonons' dispersion and dynamics nontrivial to model, but they also downconvert to acoustic phonons fast enough that that their dynamics are unimportant to capture in typical superconducting devices with \(>\mu\)m-scale dimensions.}

The set of parameters needed to integrate this physics into G4CMP for a given material is enumerated in Table~\ref{table:G4CMPParameters}. Because rigorous modeling of some of the above-mentioned physics processes in a fully anisotropic medium is nontrivial, it is often useful to simplify calculations using the ``isotropic continuum approximation,'' (ICA) in which phonons are assumed to be long enough in wavelength that anisotropy in the crystal may be ignored~\cite{srivastava1990physics}. While we (and G4CMP) use this approximation to simplify calculations of the rate coefficients for anharmonic decay and isotopic scattering, the DOS and caustics modeling account for the anisotropic crystal structure.


\begin{table*}[t!]
\caption{Parameters required for G4CMP to model phonon transport within a given solid-state material.}
\centering
\begin{tabular}{m{2.4cm}m{1.0cm}m{8.4cm}} \toprule \toprule
    {Parameter} & Units & {Description} \\ \midrule 
    C\(_{ij}\) & GPa & Second-order elastic constants \\
    \(\mu\) & GPa & Lamé constant, 2nd-order isotropic elastic constant \\
    \(\lambda\) & GPa & Lamé constant, 2nd-order isotropic elastic constant \\
    \(\beta\) & GPa & 3rd-order isotropic elastic constant \\
    \(\gamma\) & GPa & 3rd-order isotropic elastic constant \\
    \(A\) & s\(^{4}\) & Anharmonic downconversion rate coefficient \\
    \(B\) & s\(^{3}\) & Isotopic scattering rate coefficient \\
    \(F_{TT}\) & None & Fraction of \(L\rightarrow TT\) downconversion \\
    LDOS & None & Longitudinal phonons' density of states (fractional) \\
    STDOS & None & Slow transverse phonons' density of states (fractional) \\
    FTDOS & None & Fast transverse phonons' density of states (fractional) \\
    Debye Energy & THz & Debye Energy for phonon primaries \\
\bottomrule \bottomrule
\end{tabular}
\label{table:G4CMPParameters}
\end{table*}


The thrust of this section is to discuss the mathematical foundation for these physics processes and estimate the corresponding required simulation parameters for a set of substrates useful to the dark matter detection community. In doing so we also hope to provide a streamlined recipe for performing this G4CMP upgrade for any other materials later deemed of interest to the cryogenic instrumentation community.

\subsection{Anharmonic Downconversion}
\label{subsec:AnharmonicDownconversion}

Anharmonic decay of an acoustic phonon into two lower-energy acoustic phonons is phonon-polarization-dependent. Under the isotropic continuum approximation, this decay can only proceed for initial longitudinally-polarized phonons, and can only proceed via two potential decay channels: \(L\rightarrow L+T\) and \(L\rightarrow T+T\), in which the final products are a longitudinal (L) and transverse (T) phonon and two transverse phonons, respectively. Here, transverse phonons can be either on the transverse-slow phonon branch or transverse-fast phonon branch. Though beyond the scope of this work's modeling, we also note for completeness that additional decay modes, including those for transverse phonons, are possible if the ICA is relaxed~\cite{Labrot,BerkeShort,BerkeLong}.



Formally, the combined \(L\rightarrow T+T\) and \(L\rightarrow L+T\) rate of anharmonic decay within a crystal can be calculated by adding the third-order terms of the crystal's potential energy to the Hamiltonian governing the crystal's evolution. Using first-order time-dependent perturbation theory, one can derive a characteristic mode decay rate, \(\Gamma_{\mathrm{anharmonic}}\), with the following form:
\begin{equation}
\label{eq:GammaAnharmonic}
\Gamma_{\mathrm{anharmonic}}=A\nu^{5},  
\end{equation} 
where \(A\) is a constant~\cite{srivastava1990physics,Tamura_Lame_Constant} and $\nu$ is the phonon frequency. This form holds for normal (i.e. non-Umklapp) phonon processes at low temperatures (\(k_{B}T << h\nu\)). 

Calculating the constant \(A\) in Equation~\ref{eq:GammaAnharmonic} relies on knowing the second- and third-order elastic constants considered in the total potential energy of the material. Here, it is useful to apply the ICA, giving rise to a separate set of ``isotropic elastic constants:'' the Lamé constants \(\mu\) and \(\lambda\) at second order, and a set of three constants \(\alpha\), \(\beta\), and \(\gamma\) at third order. Ref.~\cite{Tamura_Lame_Constant} gives the \(L\rightarrow T+T\) and \(L\rightarrow L+T\) decay rates in terms of these parameters.

To numerically calculate the phonon anharmonic decay rate in a material of a specific crystal structure, the isotropic elastic constants \(\mu\), \(\lambda\), \(\alpha\), \(\beta\), and \(\gamma\) can be expressed using the \textit{true} second-order elastic constants \(C_{ijkl}\) and third-order elastic constants \(C_{ijklmn}\) corresponding to the material's crystal group. A general form of this parameterization is found using the development by Ref. \cite{Fedorov1968}, where Einstein notation is used:
\begin{equation} 
\label{eq:generalLameEquations}
\begin{split}
\mu & =(3C_{lklk}-C_{llkk})/30 \\[3mm]
\lambda & =(2C_{lklk}-C_{llkk})/15\\[3mm]
\alpha&=\left(8C_{iillnn}-15C_{iilnln}+8C_{inilln}\right)/105\\[3mm]
\beta &=\left(-5C_{iillnn}+19C_{iilnln}-12C_{inilln}\right)/210\\[3mm]
\gamma &=\left(2C_{iillnn}-9C_{iilnln}+9C_{inilln}\right)/210. \\[3mm]
\end{split}
\end{equation}

\begin{table*}[t!]
\caption{Expressions for the Lamé constants and third-order isotropic elastic coefficients, calculated for the two non-cubic space groups (for which such expressions could not be found in literature). The space groups for these two materials are included for completeness.}
\centering
\begin{tabular}{m{1.2cm}m{1.2cm}m{1.3cm}m{10.4cm}} \toprule \toprule
    {Material} & {Space Group} & {Parameter} & {Expression} \\ \midrule 
    \ch{Al2O3} & R$\overline{3}$c & \(\mu\) & \((2C_{11}-C_{12}-2C_{13}+C_{33}+6C_{44}+3C_{66})/15\) \\
    \cline{3-4}
    &  & \(\lambda\) & \((2C_{11}+4C_{12}+8C_{13}-4C_{44}+C_{33}-2C_{66})/15\) \\
    \cline{3-4}
    &  & \(\alpha\) & \((-17C_{111}+30C_{112}+33C_{113}+78C_{123}-57C_{133}-156C_{144}+84C_{155}-17C_{222}+C_{333}-12C_{344})/105\)\\
    \cline{3-4}
    &  & \(\beta\) & \((71C_{111}+4C_{112}-11C_{113}-68C_{123}+103C_{133}+220C_{144}-140C_{155}-55C_{222}+2C_{333}+4C_{344})/210\)\\
    \cline{3-4}
    &  & \(\gamma\) & \((-34C_{111}-24C_{112}+3C_{113}+30C_{123}-51C_{133}-144C_{144}+126C_{155}+92C_{222}+2C_{333}+18C_{344})/210\)\\
    \cline{3-4}
    \ch{CaWO4} & I4\(_{1}\)/a & \(\mu\) & \((2C_{11}-C_{12}-2C_{13}+C_{33}+6C_{44}+3C_{66})/15\)\\
    \cline{3-4}
     &  & \(\lambda\) & \((2C_{11}+4C_{12}+8C_{13}+C_{33}-4C_{44}-2C_{66})/15\)\\
     \cline{3-4}
     &  & \(\alpha\) & \((2C_{111}+18C_{112}+42C_{113}+48C_{123}-6C_{133}-60C_{144}-12C_{155}-12C_{166}+C_{333}-12C_{344}-30C_{366}+12C_{456})/105\)\\
     \cline{3-4}
     &  & \(\beta\) & \((4C_{111}+8C_{112}-7C_{113}-30C_{123}+23C_{133}+76C_{144}+4C_{155}+4C_{166}+2C_{333}+4C_{344}+38C_{366}-72C_{456})/210\) \\
     \cline{3-4}
     &  & \(\gamma\) & \((2C_{111}-3C_{112}+6C_{123}-6C_{133}-18C_{144}+9C_{155}+9C_{166}+C_{333}+9C_{344}-9C_{366}+27C_{456})/105\)\\
\bottomrule \bottomrule
\end{tabular}
\label{table:NoncubicIsotropicElasticExpressions}
\end{table*}

For a specific crystal space group these parameters can be calculated by taking advantage of symmetries present in the second- and third-order elastic constants \cite{Hearmon}. Most of the materials of interest to the dark matter search community that we consider here, including Si, Ge, GaAs, LiF and CaF\(_{2}\), belong to cubic space groups. For these, space-group-specific expressions for the second- and third-order isotropic elastic constants have been documented extensively in literature, and as a result we do not reproduce them here~\cite{Fedorov1968,Tamura_Lame_Constant,Campbell_Deem_2020_Lame}. However, a few materials of interest, including sapphire (Al\(_{2}\)O\(_{3}\)) and calcium tungstate (CaWO\(_{4}\)), have non-cubic space groups for which expressions of \(\alpha\), \(\beta\), \(\gamma\), \(\mu\), and \(\lambda\) have not yet been documented in literature (to the authors' knowledge). In Table~\ref{table:NoncubicIsotropicElasticExpressions}, we present expressions for these in terms of the materials' third-order elastic constants, using Voigt notation~\cite{Voig_Notation} for reduction from \(2n\) to \(n\) indices, where \(n\) is the order of the elastic constant.


Once these isotropic second- and third-order elastic constants are calculated for a material, the value \(A\) in Equation~\ref{eq:GammaAnharmonic} can be calculated via the treatment in Ref.~\cite{Tamura_Lame_Constant}, and used as an input to G4CMP. Results of our calculations for various materials of interest are presented in Section~\ref{subsec:AnharmonicDownconversion_Results}.

\subsection{Isotopic Scattering}
\label{subsec:IsotopicScattering}

Phonons may also be scattered on isotopic impurities in the crystal, where slight variations in atomic mass break the regularity in the crystal structure. Considering these impurities as a small perturbation to the otherwise uniform crystal Hamiltonian, one can again use perturbation theory to obtain an expression for the scattering rate of a single phonon via this channel \cite{srivastava1990physics}. This rate is proportional to the fourth power of phonon energy:
\begin{align}
\label{eq:IsotopicScattering1}
\Gamma_{\mathrm{isotopic}} =\ & B\nu^{4}\\
 =\ & \frac{4\pi^{3}\Gamma_{md}\Omega}{\langle c^{3}\rangle}\nu^{4},
\end{align}
where \(\Omega\) is the volume per atom in the crystal unit cell, \(\langle c^{3}\rangle\) is the polarization-averaged (cubed) speed of sound in the material, and \(\Gamma_{md}\) is a mass defect coefficient capturing the average deviation of the crystal from isotopic purity \cite{TamuraIsotopeCalculationGe,TamuraIsotopicCalculationsGaAs}. This expression holds true in the isotropic continuum approximation, which holds for phonons with long wavelengths relative to the interatomic spacing. The mass defect coefficient is given by \cite{Ramya,Morelli}
\begin{equation}
\label{eq:MassDefectCoefficient}
\Gamma_{md}=\frac{\left\langle \overline{\Delta M^{2}}\right\rangle }{\left\langle \overline{M}\right\rangle ^{2}},
\end{equation}
where $\left\langle \overline{\Delta M^{2}}\right\rangle$ refers to the average mass variance and $\left\langle \overline{M}\right\rangle ^{2}$ is the average mass. The average mass of each component is given by the stoichiometry-weighted average of each site average mass $\overline{M_{n}}$
\begin{align}
\label{eq:stoichAverageMass}
\left\langle \overline{M}\right\rangle & =\frac{\underset{n}{\sum}c_{n}\overline{M_{n}}}{\underset{n}{\sum}c_{n}}\\
\overline{M_{n}} & =\underset{i}{\sum}f_{i,n}M_{i,n}.
\end{align}
where the index \(n\) refers to the the \(n^{th}\) atom of the chemical formula, the index \(i\) refers to isotope \(i\) of atom \(n\), \(M_{i,n}\) refers to the atomic mass of the \(i^{th}\) isotope of atom \(n\), \(\overline{M}_{n}\) is the average isotopic mass of atom \(n\), and \(f_{i,n}\) refers to the fractional abundance of the \(i^{th}\) isotope of atom \(n\).\footnote{We note here that formally, the sum over \(i\) may be different for different atoms \(s\) in the unit cell.}
In the same way, the average mass variance is
\begin{align}
\label{eq:stoichAvgVariance}
\left\langle \overline{\triangle M^{2}}\right\rangle & =\frac{\underset{n}{\sum}c_{n}\overline{\Delta M_{n}^{2}}}{\underset{n}{\sum}c_{n}} \\
\overline{\Delta M_{n}^{2}} & =\underset{i}{\sum}f_{i,n}\left(M_{i,n}-\overline{M_{n}}\right)^{2}.
\end{align}
In this development, our use of the isotropic continuum approximation implies that Equation~\ref{eq:IsotopicScattering1} represents an average scattering rate for all acoustic phonon polarizations. While we use this definition in this work, it is worth noting that other interpretations of the mass defect coefficient have also been used \cite{Riley,Nakib}.

Once the coefficient \(B\) is calculated, it can be input into G4CMP to model isotopic scattering, where it acts as the scattering rate coefficient for all three phonon polarizations. We present the results of this calculation of \(B\) for several materials of interest in Section~\ref{subsec:IsotopicScattering_Results}.

\subsection{Density of States (DOS)}
\label{subsection:PhononDensityOfStates}

During both anharmonic downconversion processes and isotopic scattering processes (discussed in Section~\ref{subsec:IsotopicScattering}) G4CMP uses the phonon density of states to determine the randomly-drawn polarizations of phonons produced after downconversions and scatters. The LDOS, STDOS, and FTDOS parameters in Table~\ref{table:FractionalDOS} are the fractional densities of states for longitudinal, slow transverse, and fast transverse acoustic phonons, respectively, and sum to unity. 

The calculation of the DOS values for these materials is performed in two steps: computation of a 3D phonon dispersion relationship \(\omega(\vec{k})\) for a material using its lattice properties, and computation of the DOS from the phonon dispersion relationship. For each material,  we use lattice force constant information from the Materials Data Repository~\cite{Material_Project} as an input to Phonopy, a separate software used in the calculation of three-dimensional phonon dispersion relationships~\cite{phonopy-phono3py-JPCM,phonopy-phono3py-JPSJ}.\footnote{Materials files used for CaWO\(_{4}\) were from Ref.~\cite{Tanner_Plot_scattering_Phonon}.} From these dispersion curves, which are calculated at a discrete grid of $\omega(\vec{k})$ points for each branch, we compute the contribution of the DOS per phonon branch $\lambda$ at given frequency $\omega$, defined as 
\begin{equation}
g_{\lambda}(\omega)=\frac{1}{N}\underset{i}{\sum}\delta\left(\omega-\omega_{i}\right).
\end{equation} 
Here, the sum is over all phonon frequencies $\omega_{i}$ of this branch in the first Brillouin zone, and \(N\) is the total number of grid points in the first Brillouin zone. In practice, the delta function in this expression is treated as a gaussian of finite width. From this expression, we can obtain the fractional contribution at 1 THz for acoustic phonons, which is the fractional DOS input required by G4CMP. Moreover, calculating the total DOS we can obtain the maximum acoustic phonon  energy and maximum optical phonon energy (Debye Energy), the latter of which is also an input to G4CMP.

\subsection{Propagation along Caustics}
\label{subsec:PropagationAlongCaustics}

The propagation of acoustic phonons in a crystal is governed by the Green-Christoffel equation: 
\begin{equation}
\label{eq:wave_equation}
\rho\omega^{2}\epsilon_{i}=C_{ijkl}k_{j}k_{k}\epsilon_{l},
\end{equation}
where \(\omega\) is the phonon frequency, \(\epsilon_{i}\) is the phonon polarization (longitudinal, transverse fast and transverse slow), \(k\) is the phonon wavevector, \(\rho\) is the crystal's mass density, and \(C_{ijkl}\) is again the elastic constant tensor. This equation arises from considering plane monochromatic elastic wave solutions to the equations governing an elastically deformed medium.

Propagation direction is governed by the group velocity \(\vec{v}_{g} = \vec{\nabla}_{k}\omega\). As a result of an arbitrary crystal's anisotropic elastic tensor \(C_{ijkl}\), the group velocity of a phonon may not be parallel with the phonon's wavevector \(\vec{k}\) \cite{Geometric_Propagation_1,Leman_CDMS_G4CMP1}. An initially isotropic distribution of wavevector \(\vec{k}\) therefore in general gives a nonuniform distribution of group velocity. This leads to a phenomenon known as ``phonon focusing'' in a crystal, from which ``caustic'' patterns are formed (Figure~\ref{fig:SimulationRendering}). These have been measured experimentally in several experimental setups for a variety of different materials \cite{PhysRevLett.43.1424_Northrop_1,Wolfe_1998,doi:10.10520/EJC96837_Phonon_Focusing_2}.

To successfully characterize acoustic phonon propagation direction in a new material, one then only requires the second-order elastic tensor components \(C_{ijkl}\) for the new material. These elastic tensor components used in this work are available for for Al\(_{2}\)O\(_{3}\)~\cite{Second_Order_Al2O3}, GaAs~\cite{PhysRevB.32.5245_GaAs_Second_Order}, CaWO\(_{4}\)~\cite{MsallCaWO4Experiment}, CaF\(_{2}\)~\cite{BerkeShort} , and LiF~\cite{PhysRev.106.1175_LiF_Second_Order}. For simulating phonon propagation in a new material, G4CMP only requires these second-order elastic tensor components. While in practice the caustic patterns emerging from crystal anisotropies are dependent on the energy of the phonon~\cite{PhysRevB.32.5245_GaAs_Second_Order}, G4CMP does not yet include this energy dependence and instead approximates this dependence as being weak in the limit of low phonon energy.

\section{Model Results and Validation}
\label{sec:ModelResultsAndValidation}

In this section we demonstrate the above calculations as applied to Si, Ge, GaAs, Al\(_{2}\)O\(_{3}\), LiF, CaWO\(_{4}\), and CaF\(_{2}\), and where possible use existing literature measurements to do a coarse validation of our calculations and the implementation into G4CMP. These are done for all four physical properties discussed in the subsections of Section~\ref{sec:Phonon_Kinematics}: anharmonic downconversion, isotopic scattering, phonon density of states (DOS) and propagation along caustics.

\begin{table*}[t]
\caption{Anharmonic downconversion parameters for a variety of target materials of interest. The calculated Lamé parameters (\(\mu\), \(\lambda\)) and third-order isotropic elastic coefficients (\(\alpha\), \(\beta\), \(\gamma\)) (units of pressure) are used to compute the anharmonic decay rate coefficient \(A\) found in Equation~\ref{eq:GammaAnharmonic} and are included for completeness. We focus on using elastic constants that are all collected under similar cryogenic conditions. One could attempt to estimate some uncertainty on these estimates by redoing our calculations for elastic constants measured at slightly different temperatures for each material, but since it is nontrivial to find multiple such measurements, we just quote a single estimate of \(A\). For comparison, we include calculated literature values \(A_{c,l}\) and measured literature values \(A_{m,l}\) for these materials where available. Values of \(A_{c,l}\) do not always assume the same ICA used in our calculation of \(A\), and may quote a range of rates representing multiple propagation directions. \(A_{m,l}\) was difficult to find in the literature for most materials here. We were also unable to find literature values for the third-order elastic constants of CaWO\(_{4}\) for calculation of \(\alpha\), \(\beta\), and \(\gamma\), but nonetheless include CaWO\(_{4}\) as other literature calculations of \(A_{c,l}\) were found. For literature values \(A_{c,l}\) and \(A_{m,l}\), references are provided after the stated values.}
\centering
\begin{tabular}{m{1.1cm}m{0.9cm}m{1.1cm}m{1.1cm}m{1.0cm}m{1.1cm}m{0.6cm}m{1.2cm}m{2.4cm}m{1.2cm}} \toprule \toprule
    {Material} & {$\mu$} [GPa] & {$\lambda$} [GPa] & {$\alpha$} [GPa] & {$\beta$} [GPa] & {$\gamma$} [GPa] & {$F_{TT}$} & \(A\) [\(10^{-55}\)~s\(^{4}\)] & \(A_{\mathrm{c,l}}\) \, \, \, [\(10^{-55}\)~s\(^{4}\)] & \(A_{\mathrm{m,l}}\) [\(10^{-55}\)~s\(^{4}\)] \\ \midrule 
    \ch{Si}& 68.58 & 53.68 & -227.37 & -55.97 & -107.97 & 0.75 & 1.15 & 0.741 \textcolor{blue}{\cite{Tamura_Lame_Constant}} & -- \\
     \ch{Ge} & 56.6  & 33.6 & -179.3  & -60.5 & -82.1 & 0.72 & 6.8& 16.5 \textcolor{blue}{\cite{Tamura_Lame_Constant}} & -- \\
    \ch{GaAs} & 44.2  & 47.2 & -170.11  & -54.71 & -67.51 & 0.77 & 7.77 & 7.7--13.5 \textcolor{blue}{\cite{Tamura_Lame_Constant}} & -- \\
    \ch{Al2O3} & 166.24  & 139.8 & 95.13  & -27.02 & -152.8 & 0.67 & 12.7 & 1.88 \textcolor{blue}{\cite{MarisTamuraSoliton}} & -- \\ 
    \ch{LiF} & 51.51 &30.72 & -84.74 & -83.94 & -87.54 & 0.68 & 5.5 & 5.14 \textcolor{blue}{\cite{Tamura_Lame_Constant}} & -- \\
    CaF\(_{2}\) & 45.15 & 65.2 & -135.21  &-95.85  & -43.69 & 0.75 & 6.14 & 7.0--10.4 \textcolor{blue}{\cite{BerkeLong}} & 9.3 \textcolor{blue}{\cite{Baumgartner}} \\
    \ch{CaWO4} & 40.78 & 57.94 & -- & -- & -- & -- & -- & 7.9 \textcolor{blue}{\cite{WigmoreTHzPhononScattering}}-140 \textcolor{blue}{\cite{HayasakaCaWO4Anomalies}} & -- \\
\bottomrule \bottomrule
\end{tabular}
\label{table:Anharmonic_Downconversion_Results}
\end{table*}

\begin{figure}[t!]
\centering	
\includegraphics[width=\linewidth,height=5.5cm]{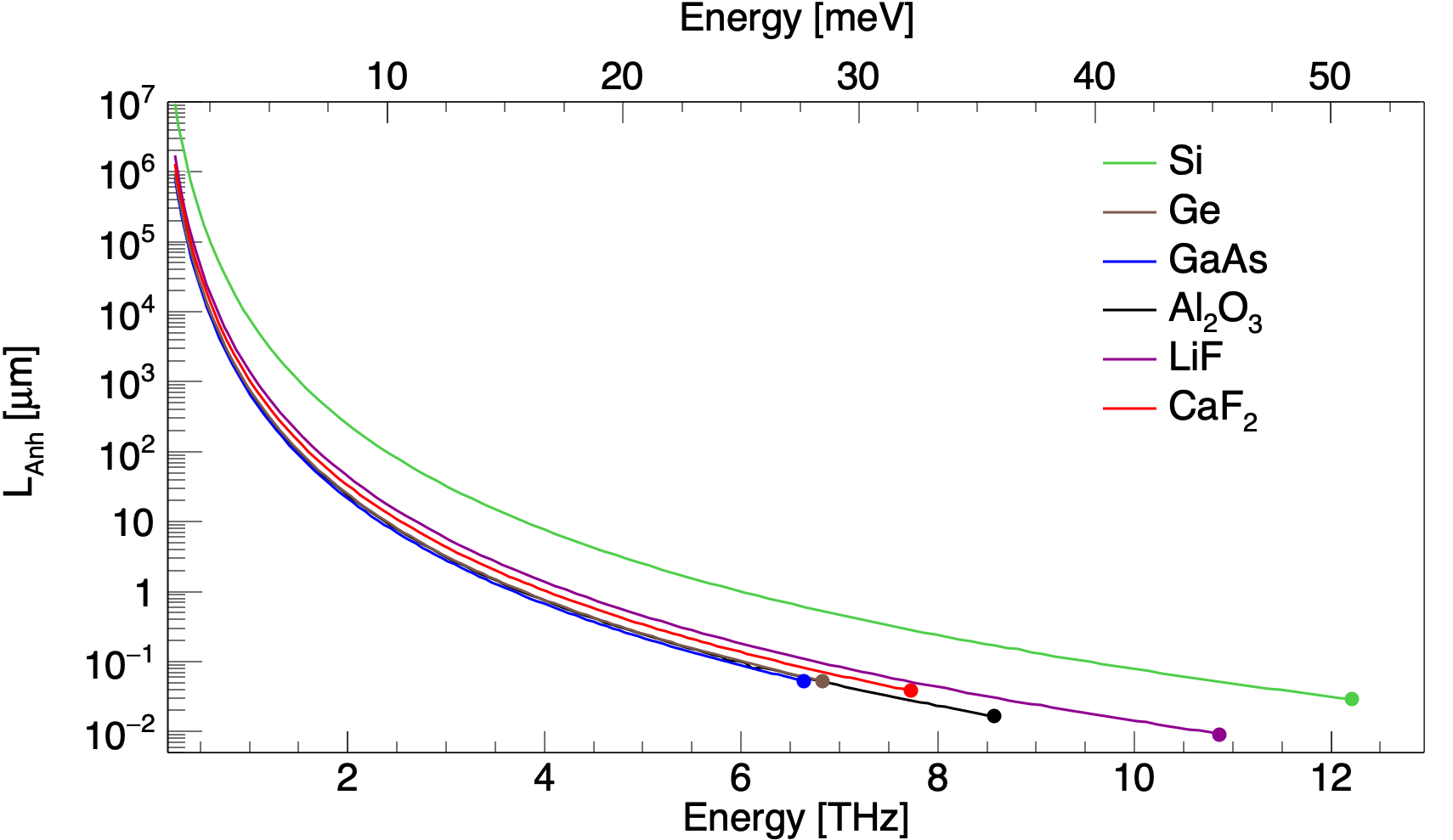}\\[3mm]
\begin{subfigure}{1.0\linewidth}
\includegraphics[width=\linewidth,height=5.5cm]{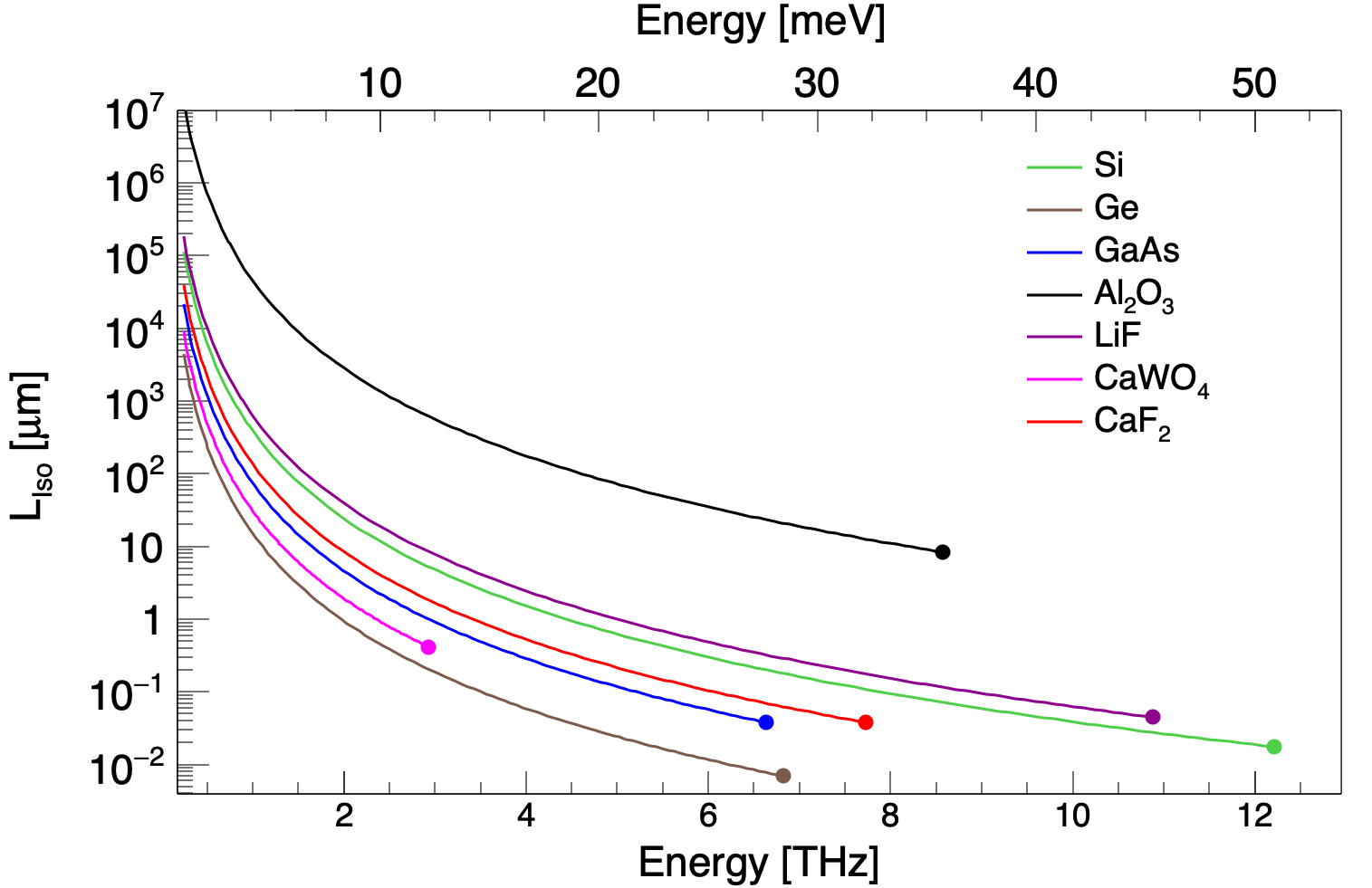}
\end{subfigure}\\
\caption{Characteristic length scales for anharmonic decay (top) and isotopic scattering (bottom) extracted from a cross-check simulation after implementing the calculated rate coefficients from Table~\ref{table:Anharmonic_Downconversion_Results}, demonstrating the \(\nu^{5}\) (top) and \(\nu^{4}\) (bottom) rate scalings. The dot at the rightmost point of each curve represents the maximum acoustic phonon energy $\omega_{A}$ extracted from table \ref{table:FractionalDOS}.}
\label{fig:DownconversionAndIsotopicScatteringPlots}
\end{figure}

\subsection{Results: Anharmonic Downconversion}
\label{subsec:AnharmonicDownconversion_Results}

Table \ref{table:Anharmonic_Downconversion_Results} presents our calculations of the second- and third-order isotropic elastic coefficients and the subsequent calculation of the parameter \(A\) in Equation~\ref{eq:GammaAnharmonic}. In these calculations, the second order elastic constants $C_{ijkl}$ used in computing \(\mu\) and \(\lambda\) as well as the third order elastic constants $C_{ijklmn}$ used in computing \(\alpha\), \(\beta\), and \(\gamma\) can be found in~\ref{sec:AppendixA}. The calculations are done assuming phonon energies of 1~THz, which are within the ``low-temperature'' approximation assumed for Equation~\ref{eq:GammaAnharmonic}. 
We also present in Table~\ref{table:Anharmonic_Downconversion_Results} two additional values, \(A_{\mathrm{c,l}}\) and \(A_{\mathrm{m,l}}\), which are calculated values for \(A\) found in literature and experimentally measured values for \(A\) found in literature, respectively. Our calculations of \(A\) agree with literature calculations of \(A_{c,l}\) to within a factor of few for most materials. There is some additional disagreement for sapphire, which we attribute to two factors: the non-isotropy assumed in the calculations in Ref~\cite{MarisTamuraSoliton}, and the steep fifth-power dependence on the assumed sound speed used to convert the results in Ref~\cite{MarisTamuraSoliton} into the form given in Equation~\ref{eq:GammaAnharmonic}. To the authors' knowledge, only one of these materials has an \(A_{m,l}\) that has been \textit{measured} directly: CaF\(_{2}\). In Ref.~\cite{Baumgartner}, this parameter is measured using an optical technique that is sensitive to phonon absorption by a doublet of an Eu\(^{2+}\) dopant in the crystal that is tunable with applied stress. We leave additional discussion on experimental validation of this parameter for Section~\ref{sec:Discussion}.

\begin{table*}[t]
\caption{Isotopic scattering rates for the set of target materials of interest. We also include some inputs to the calculation that gives those rates: the lattice cell volume \(\Omega\) and the polarization-averaged cubed sound speed \(\langle c^{3}\rangle\), as well as our calculation of the mass defect coefficient \(\Gamma_{md}\). Ranges in the \(B_{m,l}\) column correspond to measurements of different phonon polarizations. For literature values \(B_{c,l}\) and \(B_{m,l}\), references are provided after the stated values.}
\centering
\begin{tabular}{m{1.2cm}m{1.2cm}m{1.6cm}m{2.0cm}m{1.6cm}m{2.4cm}m{2.2cm}} \toprule \toprule
    {Material} & \(\Omega\) \, \, $[A^{3}]$ & \(\Gamma_{md}\) & \(\langle c^{3}\rangle\) [$10^{11}$~m$^{3}$/s$^{3}$]  & \(B\) [\(10^{-42}\)~s\(^{3}\)] & \(B_{\mathrm{c,l}}\) \, \, \, [\(10^{-42}\)~s\(^{3}\)] & \(B_{\mathrm{m,l}}\) \, \, [\(10^{-42}\)~s\(^{3}\)]\\ \midrule 
    Si & 2.0 & 2.02\(\times 10^{-4}\) & 2.13 & 2.33 & 2.42 & 2.42-2.56 \textcolor{blue}{\cite{TamuraIsotopicMeasurements1993}}  \\ 
    Ge & 2.26 & 5.88\(\times 10^{-4}\) & 0.46 & 35.4 & 36.7 \textcolor{blue}{\cite{TamuraIsotopeCalculationGe}} & -- \\
    GaAs & 2.38 & 9.16\(\times 10^{-5}\) & 0.479 & 7.22 & 7.38 \textcolor{blue}{\cite{TamuraIsotopicCalculationsGaAs}} & 5.9-29.5 \textcolor{blue}{\cite{TamuraIsotopicMeasurements1993}} \\
    Al\(_{2}\)O\(_{3}\) & 0.5 & 1.25\(\times 10^{-5}\) & 3.06 & 0.025 & -- & 0.04 \textcolor{blue}{\cite{WigmoreTHzPhononScattering}} \\
    LiF & 0.81 & 1.36\(\times 10^{-4}\) & 1.19 & 1.17 & 1.69 \textcolor{blue}{\cite{Tamura_Lame_Constant}} & -- \\
    CaF\(_{2}\) & 1.4 & 1.83\(\times 10^{-4}\)  & 0.84 & 4.83 & 9.13 \textcolor{blue}{\cite{HarringtonCaF2}} & 20.3 \textcolor{blue}{\cite{HarringtonCaF2}}\\
    CaWO\(_{4}\) & 1.3 & 2.02\(\times 10^{-4}\) & 0.22 & 15.0 & 2.4 \textcolor{blue}{\cite{WigmoreTHzPhononScattering}}-59 \textcolor{blue}{\cite{HayasakaCaWO4Anomalies}} & -- \\
\bottomrule \bottomrule
\end{tabular}
\label{table:Isotopic_Scattering_Results}
\end{table*}

With these values calculated, we also integrate them into G4CMP and run a cross-check simulation to confirm the expected interaction rate. In this simulation, one million longitudinal acoustic phonons of varying energies are generated in a large block of material and allowed to propagate with isotopic scattering processes turned off. The block is large enough that surface interactions do not affect the phonon trajectories. For every longitudinal phonon generated, the distance from creation point to subsequent decay is tabulated and inserted into a histogram corresponding to that phonon's energy. For each histogram, a characteristic decay length is tabulated, and that decay length appears as a function of energy in the top panel of Figure~\ref{fig:DownconversionAndIsotopicScatteringPlots}. In all materials tested, we find that the value of \(A\) derived from fits to these plots is consistent with the value of \(A\) used as input to G4CMP.

\subsection{Results: Isotopic Scattering}
\label{subsec:IsotopicScattering_Results}

Table \ref{table:Isotopic_Scattering_Results} presents elements of the calculation for the isotopic scattering rate coefficient \(B\) found in Equation~\ref{eq:IsotopicScattering1}. Since in the isotropic continuum approximation the isotopic scattering rate simplifies to Equation~\ref{eq:IsotopicScattering1}, the main requirement is to calculate the mass defect coefficient \(\Gamma_{md}\) for the material. For our calculations, we assume natural abundances of every atom within each material considered~\cite{Isotopes_Compositions}, which yields the values of \(\Gamma_{md}\) in the table. While isotopic enrichment of materials is possible, we present the rates from natural isotopic abundances to demonstrate a baseline strategy for how this calculation can be extended to other materials of interest for integration into G4CMP.

For isotopic scattering, we again show our own calculation of \(B\) in Table~\ref{table:Isotopic_Scattering_Results} accompanied by literature calculations \(B_{c,l}\) and literature measurements \(B_{m,l}\) along with references. Here, calculated scattering rates found in literature are generally found to be within a factor of a few of our estimates. Where there are measurements of this isotopic scattering rate, our estimates are similarly within a factor of a few of the measured literature values. Methods available in the literature have taken on three forms: thermal conductivity measurements in Ref.~\cite{HarringtonCaF2}, a phonon backscattering technique used by Ref.~\cite{WigmoreTHzPhononScattering}, and a technique in which phonons are scattered around a slot cut into the chip under test as in Ref.~\cite{TamuraIsotopicMeasurements1993}. While these measurements are more commonly available than measurements of the anharmonic decay coefficient \(A\), they are nonetheless still difficult to find for all of the materials we study. 

To test isotopic scattering's implementation into G4CMP we implement the same cross-check procedure as before, but with anharmonic decay turned off within G4CMP. The results are shown in the bottom panel of Figure~\ref{fig:DownconversionAndIsotopicScatteringPlots}. As the isotopic scattering rate is averaged over the multiple different polarizations, this represents an average characteristic length for all phonon polarizations. We again find that the characteristic decay lengths are consistent with the input rates used and a \(\nu^{4}\) dependence.

\subsection{Results: Density of States}
\label{subsec:DensityOfStates_Results}

\begin{figure}[t!]
\includegraphics[width=7.9cm,height = 6.0cm]{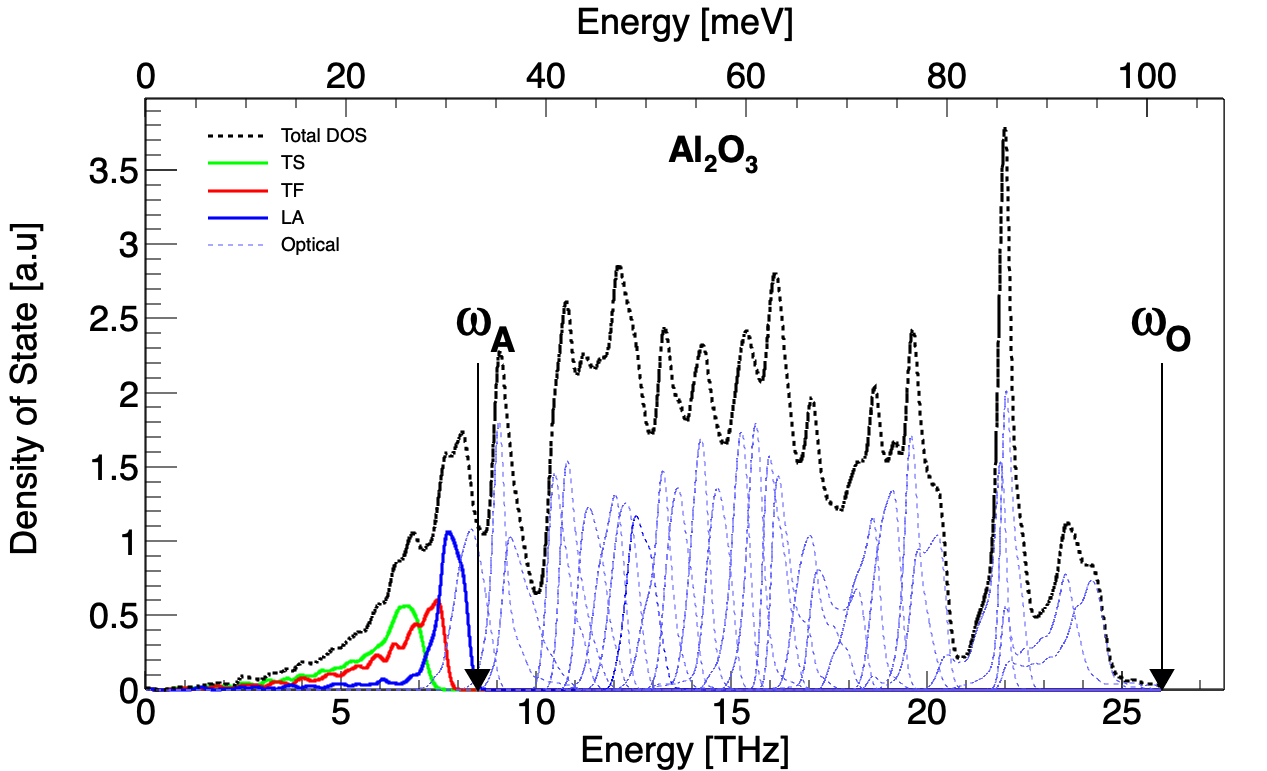}
\caption{Contribution of phonon density of states per branch, for sapphire, showing the three acoustic branches solid lines (red, green, blue) separately and showing all optical branches in the same color (dashed blue lines). The total density of states is the dashed black line. Black solid arrows are used to indicate the maximum acoustic \(\omega_{A}\) and optical \(\omega_{O}\) phonon frequencies.} 
\label{fig:DOS_Sapphire}
\end{figure}

\begin{table*}[t]
\caption{Fractional density of states STDOS:FTDOS:LDOS at 1 THz, maximum acoustic phonon frequency \(\omega_{A}\), maximum acoustic optical phonon frequency \(\omega_{O}\), and experimental maximum optical frequency \(\omega_{OE}\). The theoretical values are obtained using Phonopy~\cite{phonopy-phono3py-JPCM}. The rightmost column provides references for the \(\omega_{OE}\) column's values. Due to limited availability of Phonopy input files for Ge, the values quoted here for Ge are from Refs.~\cite{Kelsey_2023,Si_and_Ge_maximum_phonon}, and are not new calculations.}
\centering
\begin{tabular}{m{1.8cm}m{3.4cm}m{1.2cm}m{1.2cm}m{1.2cm}m{1.2cm}} \toprule \toprule
    {Material} & {STDOS:FTDOS:LDOS} & {$\omega_{A}$} [THz] &  \(\omega_{O}\) [THz]  & \(\omega_{OE}\) [THz] & Refs.  \\ \midrule 
    Si & 0.521 : 0.406 : 0.071 & 12.14 & 15.28 & 15.5 & \textcolor{blue}{\cite{Si_and_Ge_maximum_phonon} } \\ 
    Ge & 0.535 : 0.366 : 0.097  & 6.63 & 9.02 & 9.02 & \textcolor{blue}{\cite{Si_and_Ge_maximum_phonon}}   \\
    GaAs & 0.584 : 0.337 : 0.078  & 6.56 & 8.58 & 8.82 &\textcolor{blue}{\cite{GaAs_maximum_phonon_energy}}  \\
    Al\(_{2}\)O\(_{3}\) & 0.511 : 0.351 : 0.137 & 8.41 & 26.29 & 26.35 & \textcolor{blue}{\cite{Sapphire_Maximum_phonon}}   \\
    LiF & 0.5156 : 0.386 : 0.0983  & 10.73 & 18.19 & 19.7 & \textcolor{blue}{\cite{LiF_maximum_phonon}}  \\
    CaF\(_{2}\) & 0.599 : 0.318 : 0.0819  & 7.55  & 13.65 & 13.67 & \textcolor{blue}{\cite{CaF2_maximum_phonon}}\\
    CaWO\(_{4}\) & 0.475 : 0.408 : 0.115 & 2.72 & 25.94 & 26.07 & \textcolor{blue}{\cite{CaWO4_maximum_phonon}} \\
\bottomrule \bottomrule
\end{tabular}
\label{table:FractionalDOS}
\end{table*}

For each of the crystal structures we study in this work, we compute phonon dispersion curves for the various phonon branches, and use these to construct branch-specific density of states curves as a function of energy (Figure~\ref{fig:DOS_Sapphire}). Fractional density of states contributions are then computed for acoustic longitudinal, fast transverse, and slow transverse phonons at 1~THz, with the results shown in Table \ref{table:FractionalDOS}. These values are used as inputs to G4CMP. Energy-dependent DOS curves are shown for additional materials in \ref{sec:AppendixB}.

\subsection{Results: Propagation along Caustics}
\label{subsec:PropagationAlongCaustics_Results}

\begin{figure}[t!]
\includegraphics[width=7.5cm]{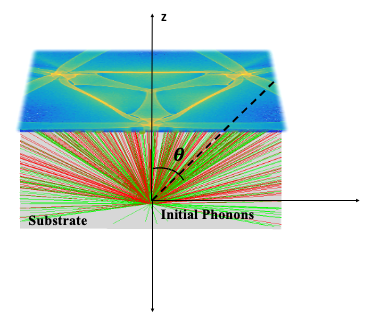}
\caption{A rendering of the simulation to test G4CMP's propagation along caustics. The gray box is the substrate (sapphire, in this image), the red lines are transverse fast phonon trajectories, the green lines are the transverse slow phonon trajectories, and $\theta$ is the angular scan range. The blue rectangle on the top surface of the substrate is the phonon collection plane, where brighter colors indicate more densely concentrated phonon impacts. In this visualization, phonons were generated at the bottom of the substrate.}
\label{fig:SimulationRendering}
\end{figure}

\begin{figure}[t!]
\centering
\begin{subfigure}{1.0\linewidth}
\includegraphics[width=\linewidth]{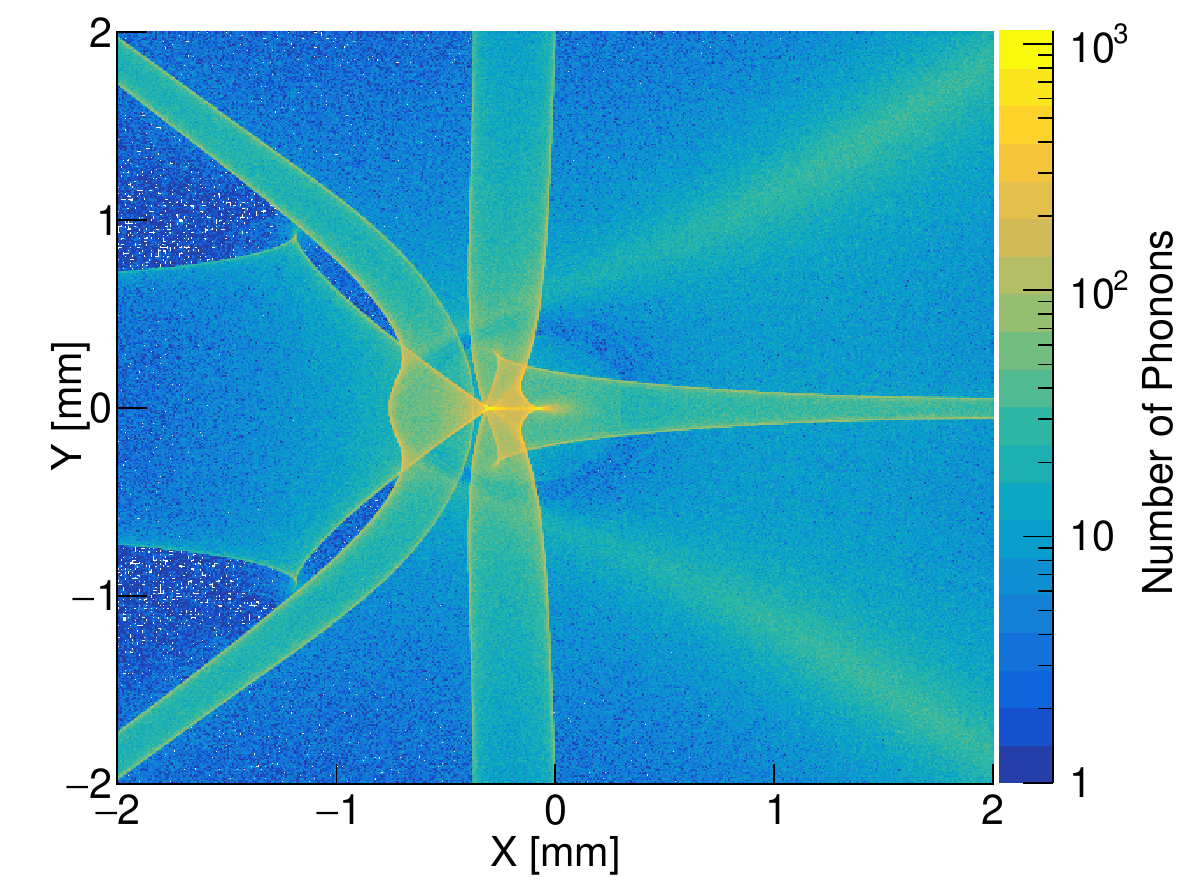}
\end{subfigure}\\
\begin{subfigure}{0.8\linewidth}
\includegraphics[width=\linewidth]{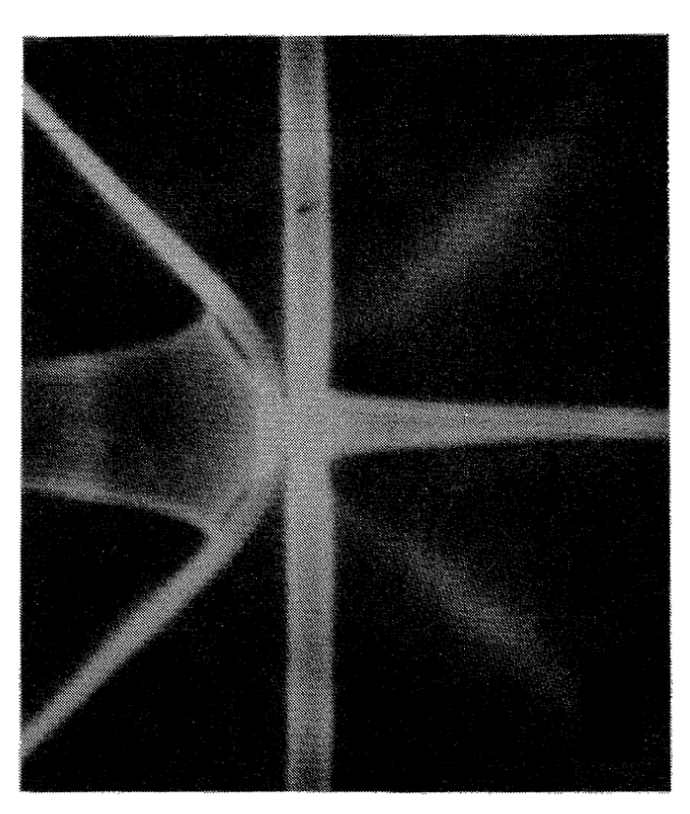}
\end{subfigure}\\
\caption{Top: phonon caustic image for sapphire (Al\(_{2}\)O\(_{3}\)) obtained from G4CMP. The crystal direction [1\(\bar{1}\)02] is at the center of the pattern, oriented out-of-page. Bright regions indicate directions of high phonon flux. Bottom: phonon caustic image for sapphire measured at 1.6~K in Ref.~\cite{Ballistic_phonons_and_elastic_Constants}, where the crystal direction [1\(\bar{1}\)02] is out of page and horizontal scan range is $\pm 32^{\circ}$. This angular scan range (and that in all subsequent caustic images) matches that in the above simulations.}
\label{fig:SapphireCausticsSim_1102}
\end{figure}

\begin{figure}[h!]
\centering
\begin{subfigure}{1.1\linewidth}
\includegraphics[width=\linewidth]{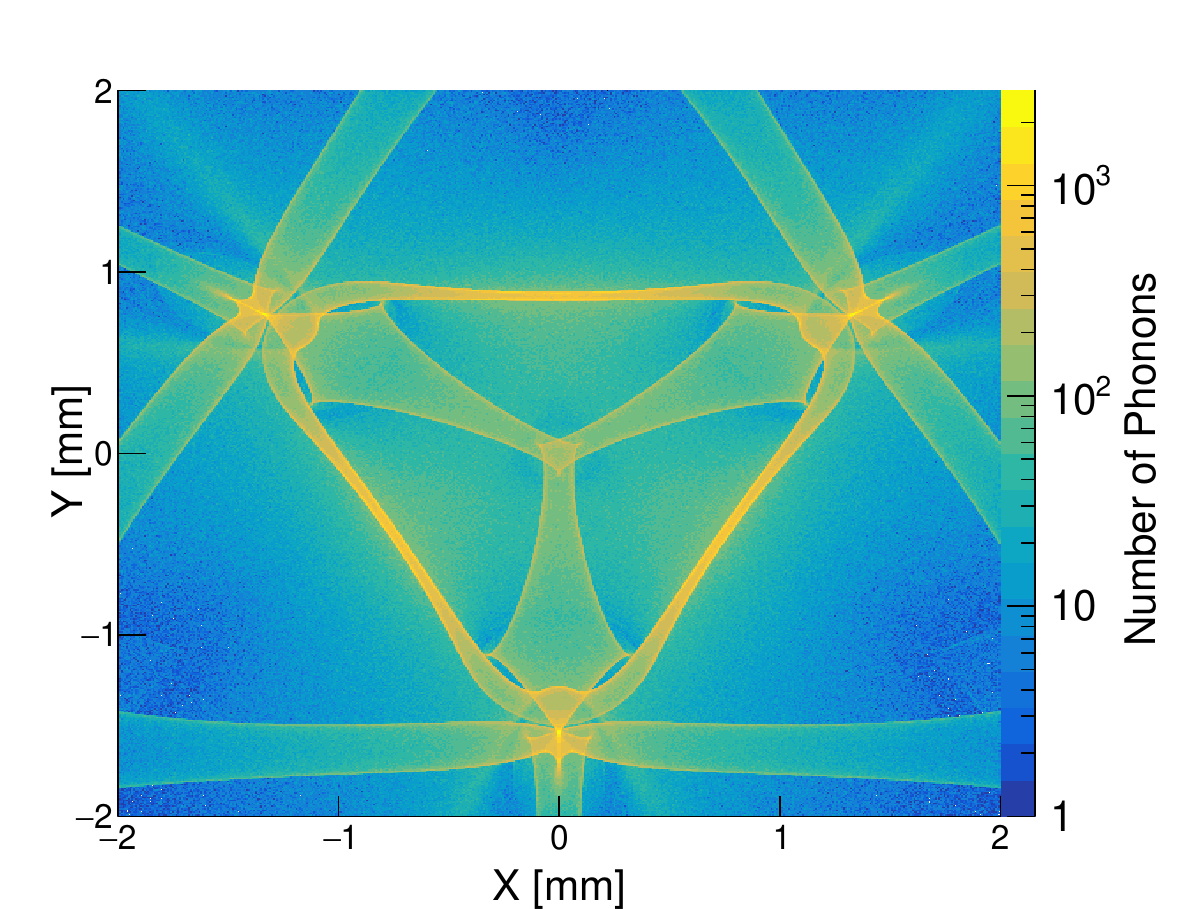}
\end{subfigure}\\
\begin{subfigure}{\linewidth}
\includegraphics[width=\linewidth]{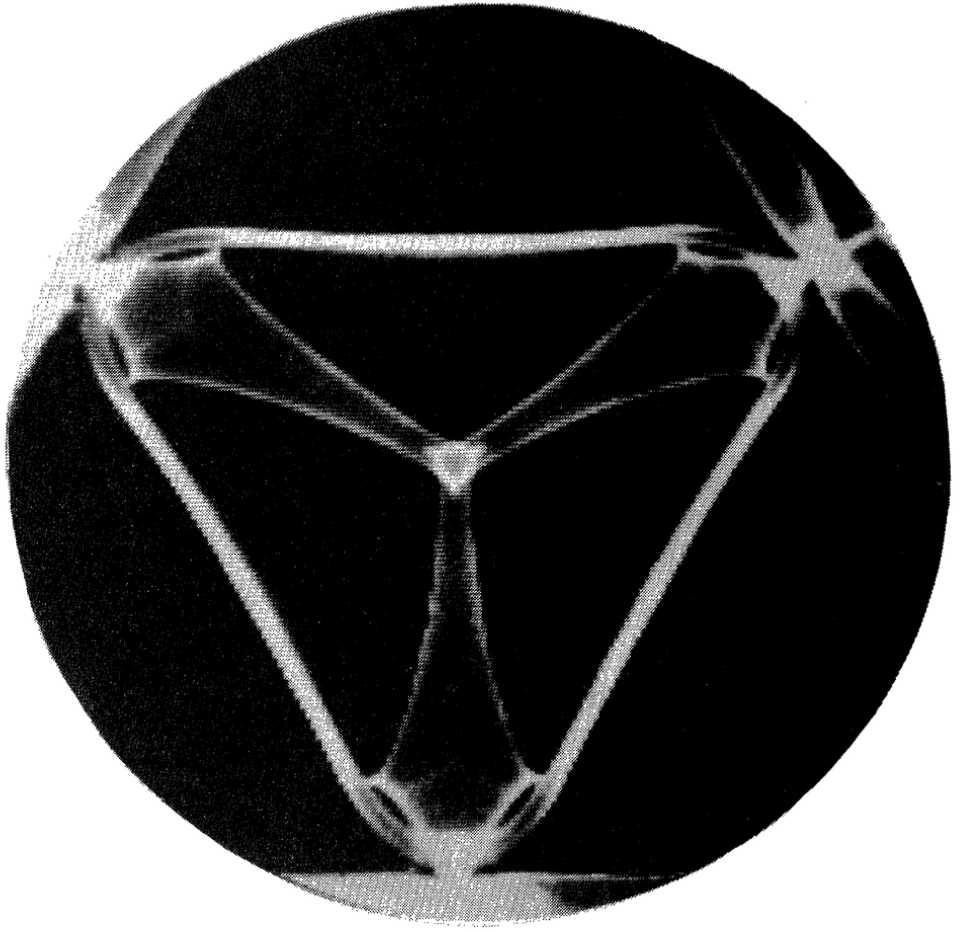}	
\end{subfigure}\\
\caption{Top: phonon caustic image for sapphire (Al\(_{2}\)O\(_{3}\)) obtained from G4CMP. The crystal direction [0010] is at the center of the pattern, oriented out-of-page. Bottom: phonon caustic image for sapphire measured at 1.6~K in Ref.~\cite{Ballistic_phonons_and_elastic_Constants}, where the crystal direction [0010] is out of page and horizontal scan range is $\pm 58^{\circ}$.}	
\label{fig:SapphireCausticsSim_0010}
\end{figure}

\begin{figure}[t!]
\centering
\begin{subfigure}{1.0\linewidth}
\includegraphics[width=\linewidth,height=6cm]{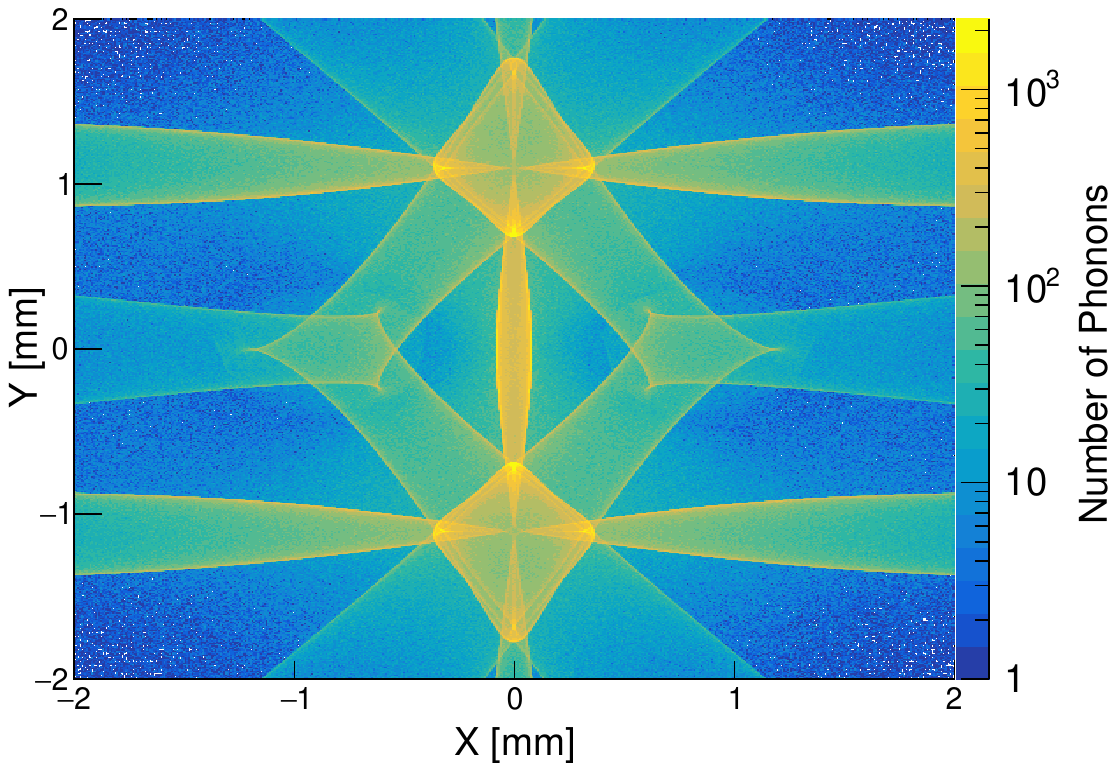}		
\end{subfigure}\\
\begin{subfigure}{0.8\linewidth}
\includegraphics[width=\linewidth,height=6cm]
{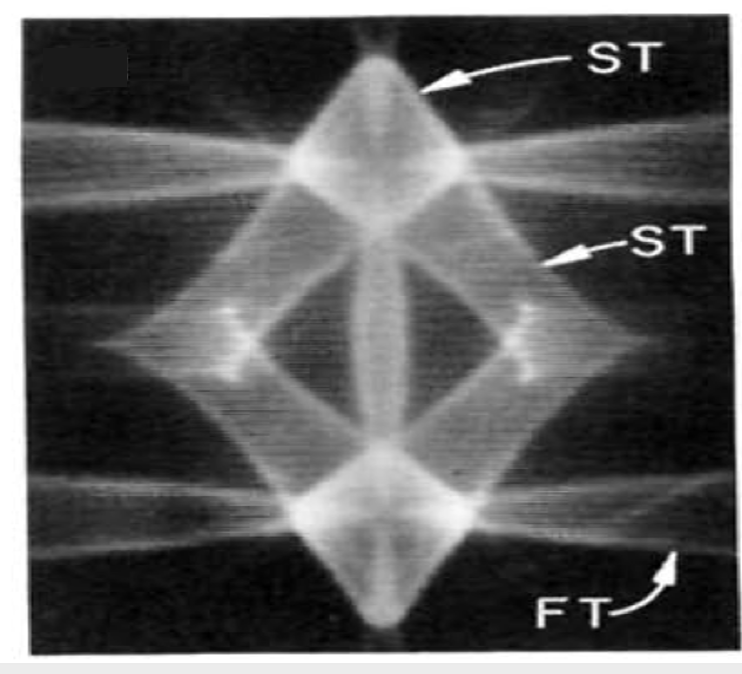}
\end{subfigure}\\
\caption{Top: phonon caustic image for GaAs obtained from G4CMP. The crystal direction [110] is at the center of the pattern, oriented out-of-page. Bottom: phonon caustic image for GaAs measured at 1.8~K in Ref.~\cite{Experimental_GaAs}, where the crystal direction [110] is out of page and horizontal scan range is $\pm 59^{\circ}$.}
\label{fig:PhononCausticGaAs}
\end{figure}

\begin{figure}[t!]
\centering
\begin{subfigure}{1.0\linewidth}
\includegraphics[width=\linewidth,height=6cm]{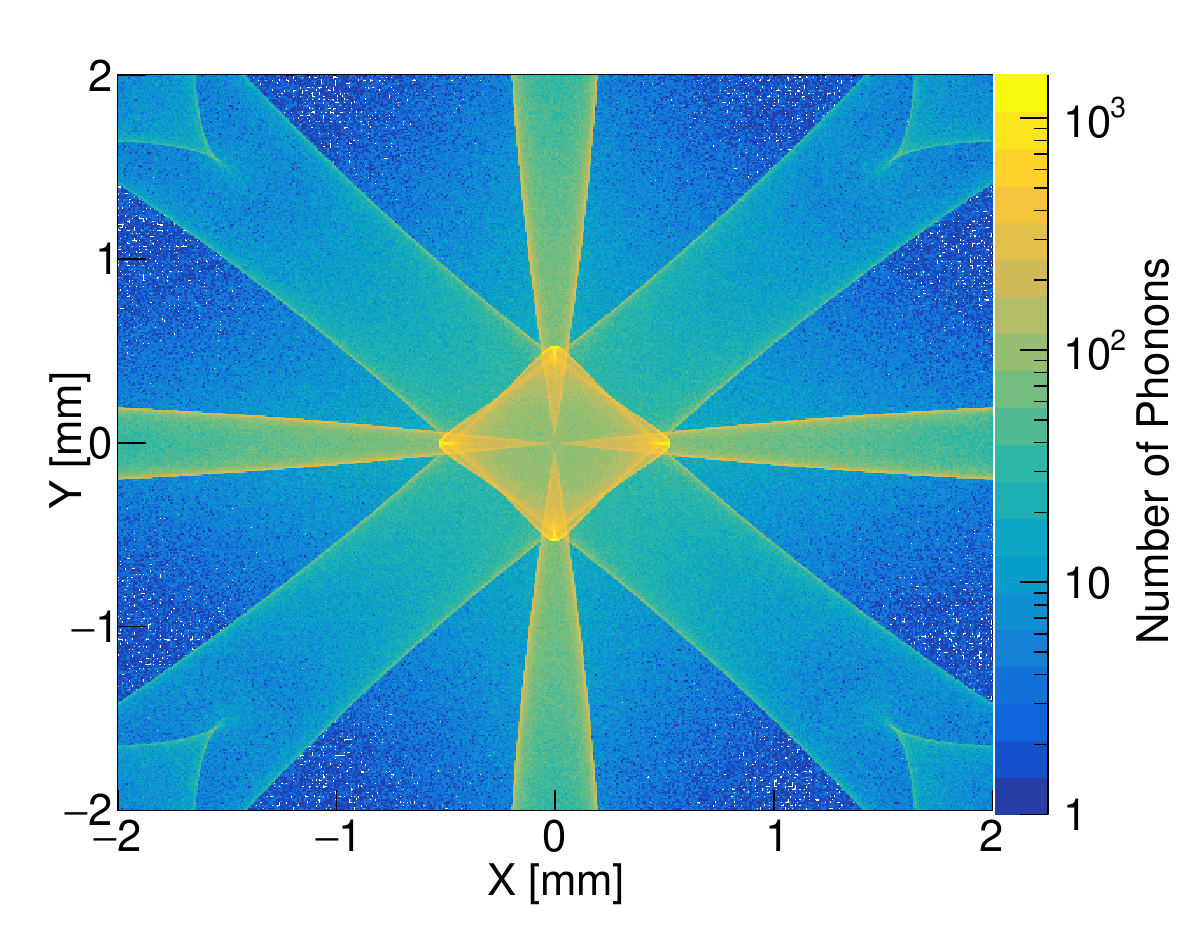}
\end{subfigure}\\
\begin{subfigure}{0.75\linewidth}
\includegraphics[width=\linewidth,height=6cm]{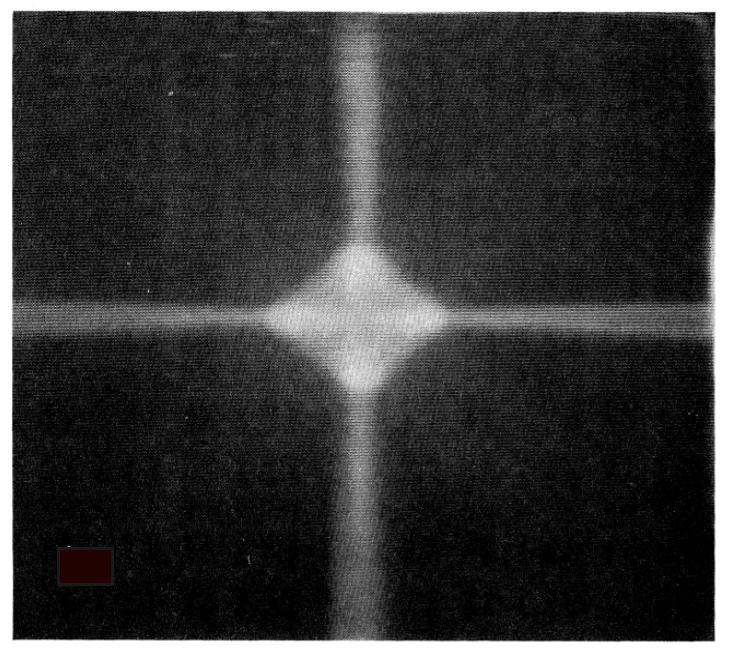}
\end{subfigure}\\
\caption{Top: phonon caustic image for LiF obtained from G4CMP. The crystal direction [100] is at the center of the pattern, oriented out-of-page.  Bottom: phonon caustic image for LiF measured at 2.2~K in Ref.~\cite{NorthropLiFPRL}, where the crystal direction [100] is out of page and horizontal scan range is $\pm 40 ^{\circ}$.}
\label{fig:PhononCausticLiF}
\end{figure}

\begin{figure}[h!]
\centering
\begin{subfigure}{1.0\linewidth}
\includegraphics[width=\linewidth,height=6cm]{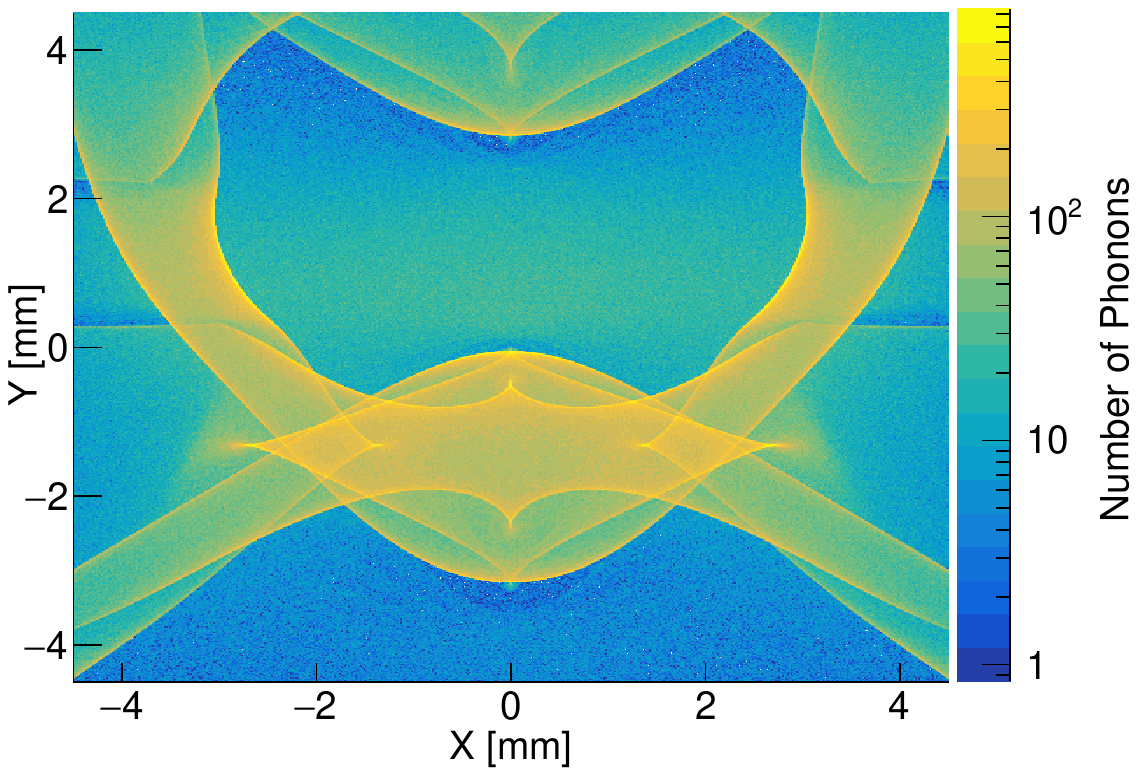}		
\end{subfigure}\\
\begin{subfigure}{0.8\linewidth}
\includegraphics[width=\linewidth,height=5.6cm]{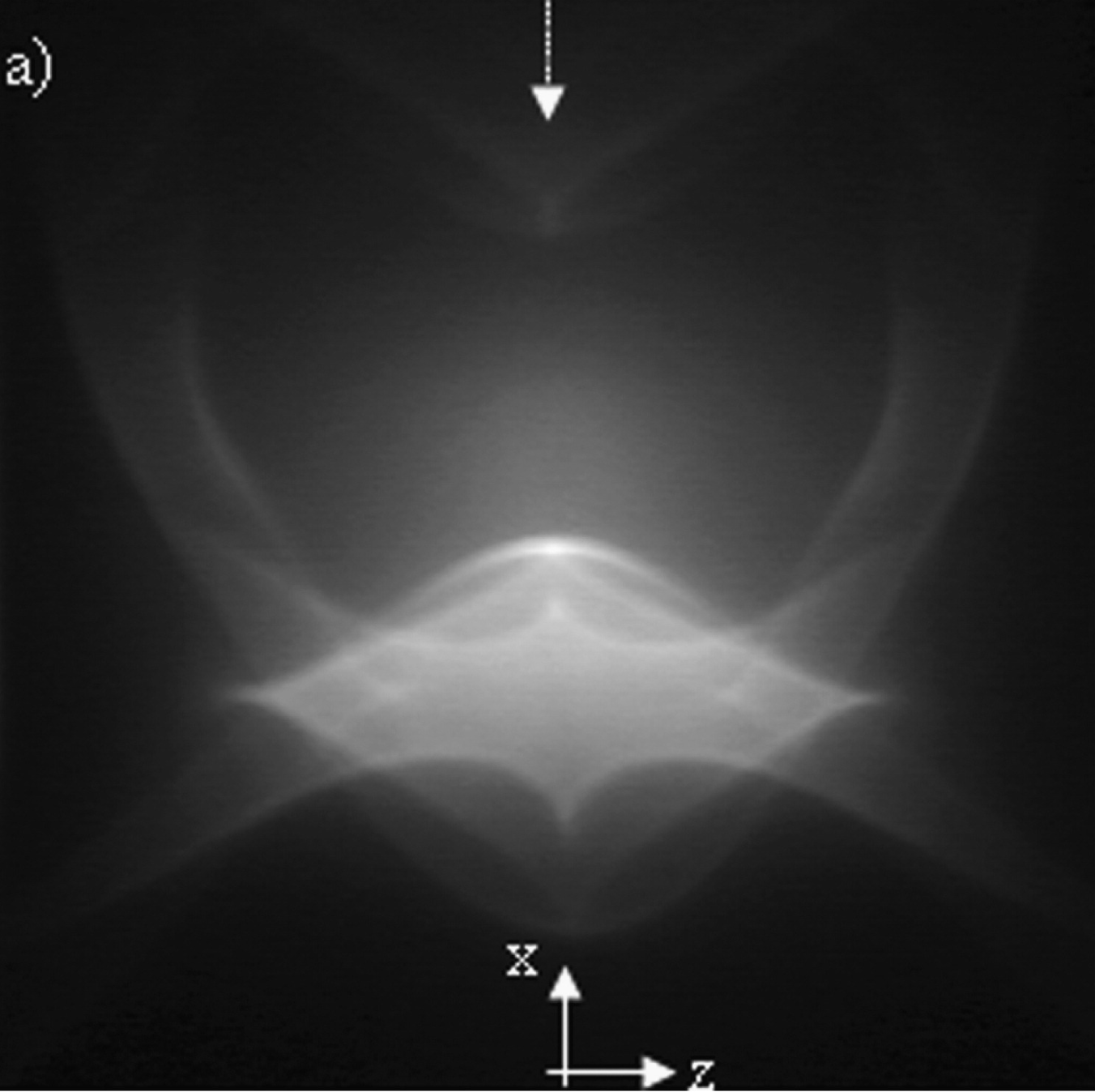}		
\end{subfigure}\\
\caption{Top: phonon caustic image for CaWO\(_{4}\) obtained from G4CMP. The crystal direction [010] is at the center of the pattern, oriented out-of-page. Bottom: phonon caustic image for CaWO\(_{4}\) measured below 2~K in Ref.~\cite{MsallCaWO4Experiment}, where the crystal direction [010] is out of page and horizontal scan range is $\pm 56.3^{\circ}$. In the simulation used to produce the top image, the crystal has been laterally extended to 9.1~mm square and 3~mm thick to enable a match to the scan range from the (bottom) experimental data.}
\label{fig:PhononCausticCaWO4}
\end{figure}

\begin{figure}[h!]
\centering
\begin{subfigure}{1.0\linewidth}
\includegraphics[width=\linewidth,height=6cm]{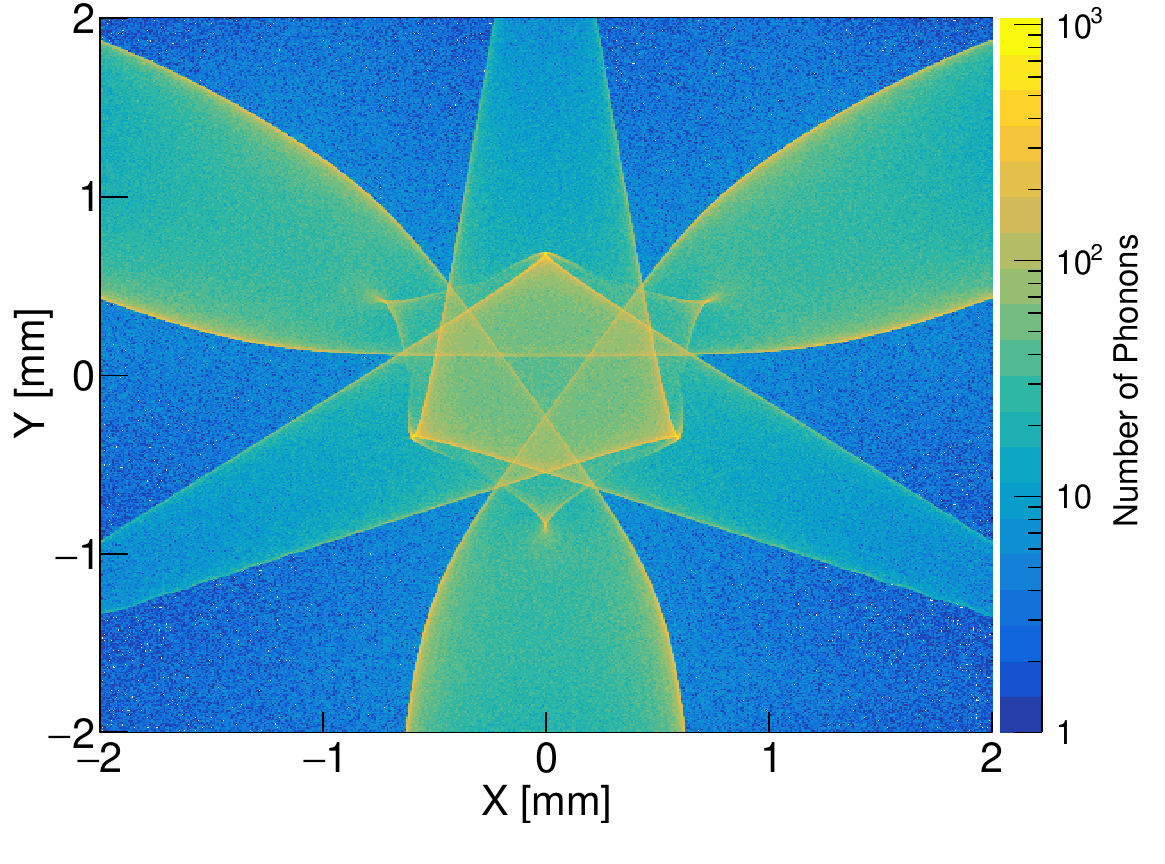}		
\end{subfigure}\\
\begin{subfigure}{0.8\linewidth}
\includegraphics[width=\linewidth,height=6cm]{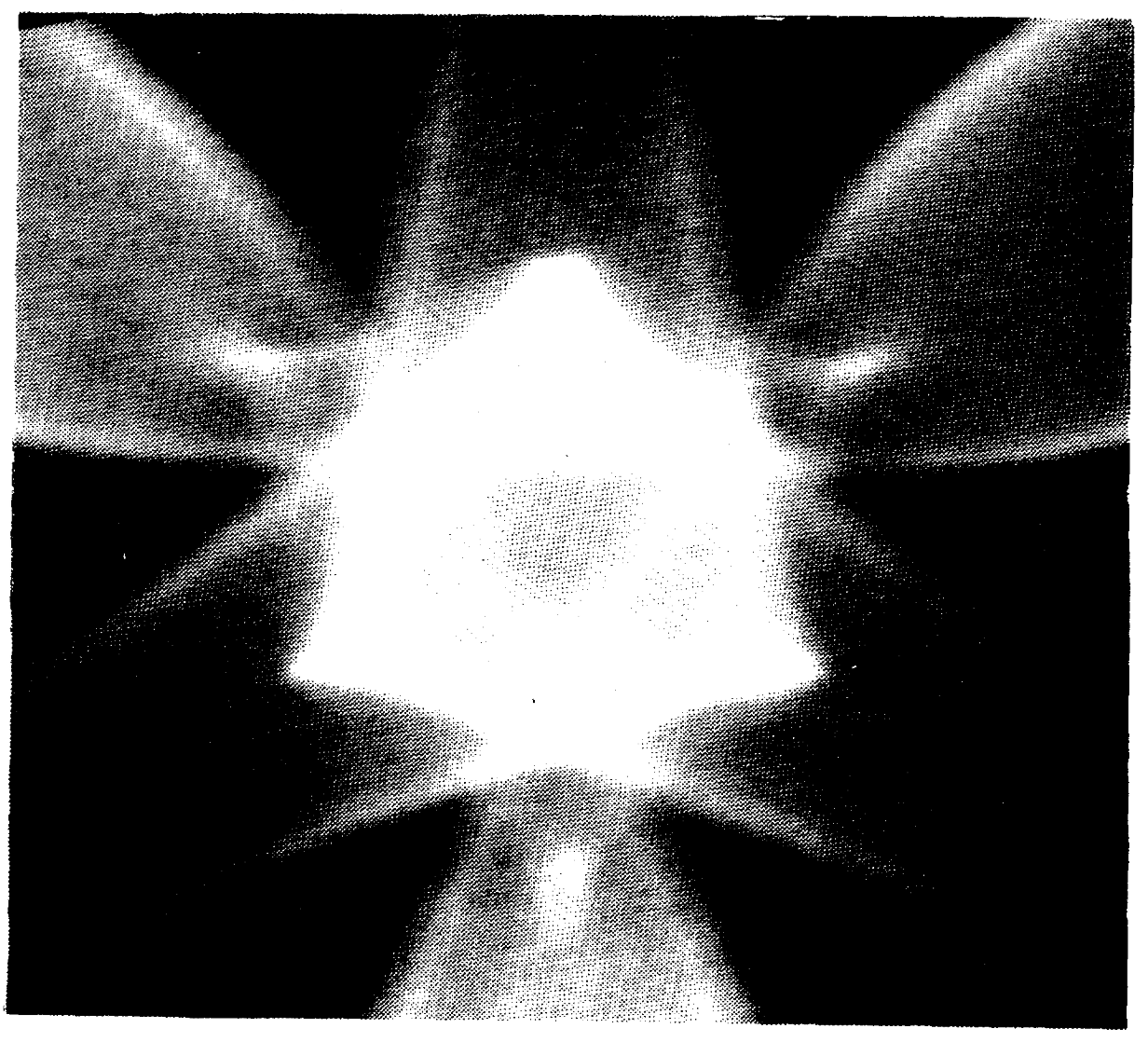}		
\end{subfigure}\\
\caption{Top: phonon caustic image for CaF\(_{2}\) obtained from G4CMP. The crystal direction [111] is at the center of the pattern, oriented out-of-page. Bottom: phonon caustic image for CaF\(_{2}\) measured in Ref.~\cite{HurleyCausticsCaF2}, where the crystal direction [111] is out of page and horizontal scan range is $\pm 23^{\circ}$. No measurement temperature was reported.}	
\label{fig:PhononCausticCaF2}
\end{figure}

To demonstrate successful implementation of the physical processes responsible for phonon propagation along caustics, we compare simulated caustic patterns against experimental heat-pulse caustic measurements. For each material, we create a simple G4CMP geometry consisting of a 4mm cube of substrate with its +\textit{z} face acting as a phonon collection plane (similar to Figure~\ref{fig:SimulationRendering}). Anharmonic downconversion and isotopic scattering are turned off to have ballistic phonons. 40 million low-energy phonons (at 1 THz) are simulated isotropically from a point at \(x=0\), \(y=0\), and a \textit{z} chosen to ensure that the collecting plane subtends the same angular scan range \(\theta\) as those reported in experimental caustics probes. The locations of phonons landing on the collection plane are recorded, and histogrammed. The ratio of longitudinal to transverse-fast to transverse-slow polarizations used here is 0:1:1. This ratio is used primarily for comparison with the experimental data (see below), which often uses arrival-time-based gates to select the transverse phonon signal and cut away the longitudinal signal.\footnote{It is also worth noting that the longitudinal phonon signal commonly shows less pronounced caustic structures than transverse phonon signals in a variety of materials, often displaying only a central ``hotspot'' of detected phonons in such images.} The top panel in Figure~\ref{fig:SapphireCausticsSim_1102} (Figure~\ref{fig:SapphireCausticsSim_0010}) shows the resulting caustic patterns for sapphire, with the crystal direction [1\(\bar{1}\)02] ([0010]) oriented out of the page. We also show simulated caustics for GaAs (Figure~\ref{fig:PhononCausticGaAs}), LiF (Figure~\ref{fig:PhononCausticLiF}), CaWO\(_{4}\) (Figure~\ref{fig:PhononCausticCaWO4}), and CaF\(_{2}\) (Figure~\ref{fig:PhononCausticCaF2}).

We validate our simulations against experimental caustics measurements from literature, shown in the bottom panels of Figures~\ref{fig:SapphireCausticsSim_1102}-\ref{fig:PhononCausticCaF2}. Phonon caustics imaging has been achieved in a variety of materials, including all of the materials shown in Table~\ref{table:Anharmonic_Downconversion_Results}. Such experiments typically proceed as follows. A crystal substrate under test is outfitted with a small, single bolometer (area \(\simeq 50\upmu \mathrm{m} \times 50\upmu \mathrm{m}\)) on one side and a metal film on the other, and is then cooled to cryogenic temperatures (few~K scale or below). A phonon burst is then produced at cryogenic temperatures by passing a short burst of current through the metal film or by illuminating the metal film locally with laser light. In this process, several non-spherical shells of longitudinal and transverse phonons are produced locally. Of these locally-created phonons, only those traveling in the direction of the bolometer are detected. This process is repeated for multiple source locations. While this experimental data-taking scheme is geometrically different from our simulation, functionally the mappings are identical: the pattern of phonons radiated from a point onto a surface (collecting plane) should be the same as the pattern of phonons radiated from an extended surface onto a point (bolometer). This enables direct comparisons between the simulated and experimentally-acquired images of these phonon caustics. In the comparisons we present, exact normalization of the different polarization components is not known for many of the experimental images, but in most\footnote{NB: Ref.~\cite{MsallCaWO4Experiment} does not specify if these are time-gated to select only transverse phonons.} cases the phonon images have been time-gated to select only transverse phonons. We therefore attempt to coarsely match this weighting of polarizations by using a 0:1:1 ratio for L:TF:TS phonons, but acknowledge that this may not yield a perfect match to the relative amplitudes of the measured caustic signals. 

Even despite this uncertainty in the polarization weighting, the shapes of the resulting caustic images constructed with G4CMP and the shapes of the corresponding experimental images agree well, given the identical lateral scan ranges. This implies successful integration of these new materials' phonon transport properties into G4CMP. Direct quantitative comparisons of the brightness of different regions is challenging given the format of the literature data and the unknown degree to which the experimental response displays a saturation in intensity. For this reason, we focus on shape comparisons.\footnote{This also contributes to our decision to display our plots with a logarithmic intensity (z-)axis: even though they are less comparable to the literature heat-pulse images, such logarithmic plots give more insight into the caustic structures than what is available in the sometimes-saturated experimental images.}

It is worth noting that many of the studies from which we've obtained the experimental images also perform caustic simulations which show good agreement with their experimental results. However, to the authors' knowledge, these simulations have largely been done with separate custom simulation software dedicated to exploring caustics. This current work is not only a demonstration of a framework that we can use to reproduce similar results within a single public-facing software package, but also that we can do so within a package whose primary goal is an application of this physics to understanding low-energy detector response.

\section{Discussion}
\label{sec:Discussion}

The primary goal of this work has been to present and demonstrate a streamlined framework for integrating new materials into the G4CMP low-energy simulation software. The striking agreement between simulated and measured caustic diagrams for sapphire, LiF, GaAs, CaWO\(_{4}\), and CaF\(_{2}\) illustrate the general success of this implementation and the relatively small set of required parameters needed for modeling this phonon transport within a publicly available software package. Moreover, the similar simplicity in input parameters for modeling anharmonic decay and isotopic scattering continues to paint an encouraging picture of accessibility for efforts to model phonon transport in future novel detector materials. 

Given this modeling success, three natural topics of future interest are how one can further improve G4CMP's modeling capabilities, what work needs to be done to validate these models, and how this framework plays a role in particle detection and QIS applications.

From this work, we identify a few areas where future effort could be reasonably made to further improve realism in G4CMP's phonon modeling. First, the calculations here could be redone without using the isotropic continuum approximation in modeling downconversion and isotopic scattering. In G4CMP, the rates of these processes are agnostic to the direction of the decaying/scattering phonon. Ref.~\cite{BerkeLong} demonstrates not only that the rates of phonon downconversion can vary by factors of a few depending on the direction of propagation, but also that in a rigorously-modeled anisotropic scenario, more decay modes are possible than the \(L\rightarrow T+T\) and \(L\rightarrow L+T\) that G4CMP currently models: both transverse fast and transverse slow phonons may undergo anharmonic decay. As G4CMP preserves the phonon direction and polarization information step-by-step, a natural extension of this work may therefore be to implement this direction-dependence and more complex polarization dependence of rate. Furthermore, while phonon scattering on isotopic impurities is handled as being isotropic in momentum (k-)space according to G4CMP, this is not universally true: experiments in Ref~\cite{TamuraIsotopicMeasurements1993} indicate that there is also a degree of anisotropy to be expected in the phonon scattering process, beyond that incurred from just propagation along caustics from the scattering site. Second, an improvement to G4CMP may include accounting for dispersion's impact on the phonon caustics, which currently do not (but should) change as a function of phonon energy~\cite{PhysRevB.32.5245_GaAs_Second_Order}. This energy dependence may also be responsible for small degrees of disagreement observed between simulated and measured caustics patterns in Figures~\ref{fig:SapphireCausticsSim_1102}-\ref{fig:PhononCausticCaF2}. Other improvements to G4CMP include modeling electron/hole transport and processes in these newly-added materials, adding photon interactions with the substrate, and developing functionality for quanta to traverse boundaries between different crystals (such as the interface between a substrate and its superconducting film). While a thorough discussion of these efforts is beyond the scope of this work, they will be the topic of follow-up studies.

This work has also highlighted strategies used to validate phonon transport models in our expanded set of target materials, as well as the fact that measurements are still needed to perform this validation for many of the materials considered. Measurements of the anharmonic decay rate coefficient, \(A_{m,l}\), are largely absent: we were only able to find one direct measurement \(A_{m,l}\) for CaF\(_{2}\). This direct measurement relies on doping a crystal with Eu\(^{2+}\) atoms and taking advantage of the stress-dependence of optical transitions in that atom. Employing this strategy in an arbitrary material clearly runs into prohibitive challenges when the material bandgap is lower than the energy of the optical transitions accessible, which may help explain why \(A_{m,l}\) has not been measured (to the authors' knowledge) for any of the optically opaque materials studied here. This strategy may be viable for directly measuring anharmonic decay rates in other optically clear materials, but it is unclear how broadly applicable it may be in validating estimates of \(A\) for arbitrary materials. Another possible (albeit less direct) means of measuring this may follow the heat-pulse strategy employed in Ref.~\cite{WigmoreTHzPhononScattering}. Here, heat pulse transmission measurements are taken at cryogenic temperatures using a bolometer, and for sufficiently large heater powers, they are (plausibly) affected by anharmonic downconversion in the material. While such measurements are also affected by isotopic scattering and would have to simultaneously fit for both \(A\) and \(B\), they nonetheless may grant some sensitivity to directly measuring downconversion rates. The isotopic scattering rate \(B\) seems more straightforward to measure than \(A\), which may be partly due to the more diverse set of measurement techniques cited in Table~\ref{table:Isotopic_Scattering_Results}. However, we were still not able to find a measured isotopic scattering rate coefficient \(B_{m,l}\) for all materials. Overall, these observations illustrate that most materials are missing measurements of fundamental phonon transport parameters needed for validation of transport models in our expanded set of materials.

Lastly, it is worth discussing how the tools developed here strengthen particle detection and QIS applications. While we used this framework to model several new materials, it can be readily extended in kind to arbitrary solid-state materials that are well-motivated from low-threshold sensing and QIS standpoints. Diamond, SiO\(_{2}\) and AlN are particularly well-motivated dark matter targets as they are advantageous for looking for single-phonon excitations produced from a DM interaction through a hadrophilic scalar mediator~\cite{Tanner_Plot_scattering_Phonon}. Other materials like SiC are also well-motivated as DM targets, as they may enable sensitive probes of a broad set of DM models~\cite{SiCDetectorsForDM}. More complex crystals with low electronic bandgaps such as Eu\(_{5}\)In\(_{2}\)Sb\(_{6}\) are also of current interest for their lowered threshold (relative to Si and Ge) for charge readout~\cite{WatkinsTaup2023}. If those are also used in an NTL phonon-amplification and sensing scheme, being able to understand phonon propagation within them will also be critical for detector response modeling. Other materials such as Li\(_{2}\)MoO\(_{4}\) and TeO\(_{2}\) might also modeled to the benefit of the neutrinoless double beta decay communities, where detector energy resolution, not detector energy threshold, is commonly prioritized~\cite{LMO,TeO2}. From a QIS perspective, Si and Al\(_{2}\)O\(_{3}\) are the most common substrates used, and are now both modeled in G4CMP. However, superconductors such as aluminum, niobium, and tantalum are all commonly used in ground plane and circuit design. Coupling this framework (applied to those materials) with an expansion of G4CMP's ability to transport phonons between different crystals will enable rigorous characterization of microphysics in the superconducting layer, which is important for modeling quasiparticle-induced qubit errors. Though it is challenging to discuss and analyze all well-motivated materials candidates for sensing and QIS, the framework presented here will nonetheless enable more efficient modeling of these materials in G4CMP. 

The results of this work may provide additional use in applications where particle impacts are expected to be spatially localized within a device. In such a scenario, these phonon models (and notably caustic patterns) may enable one to place sensors to maximize the chances of observing the emergent phonon pulses created from the impact point~\cite{Kelsey_2023,KIPM_Dylan}. As more materials' phonon transport properties are integrated into G4CMP, such strategies may be more easily used to improve a device's detection efficiency. Overall, this framework's general assistance in understanding a material's microphysical phonon response provides the foundation for modeling signal and background responses in particle physics experiments and for modeling phonon-induced correlated errors in superconducting qubits. 

\section{Conclusion}
\label{sec:Conclusions}

In this work, we have presented a condensed framework for adding phonon transport modeling into the G4CMP simulation package for new materials of interest to the dark matter direct detection and QIS communities. We have used this framework to explore phonon transport parameters needed by G4CMP for seven materials, five of which have not yet been integrated into a comprehensive low-energy simulation package: GaAs, Al\(_{2}\)O\(_{3}\), LiF, CaWO\(_{4}\), and CaF\(_{2}\). Moreover, we have accompanied this exploration with a literature search of calculated and measured phonon transport parameters for those materials, and have found that the integration of our calculations into G4CMP gives phonon propagation that is largely consistent with calculated and measured decay rates, scattering rates, phonon density of states, and phonon caustics in those materials. As measurements for all materials were not found in this search, it is also apparent that there is further need for experimental work to help validate these phonon transport parameters for many materials of interest.


\section{Acknowledgements}
The authors would like to thank Tanner Trickle for providing insights and suggesting strategies for calculating the density of states for various materials. We also thank Joyce Christiansen-Salameh for providing insight into how to access the materials information to be used with Phonopy. We thank the rest of the Fermilab QSC group for freeing up person-hours for the authors to develop this work. This manuscript has been authored by Fermi Research Alliance, LLC under Contract No. DE-AC02-07CH11359 with the U.S. Department of Energy, Office of Science, Office of High Energy Physics. This work was supported by the U.S. Department of Energy, Office of Science, National Quantum Information Science Research Centers, Quantum Science Center, the U.S. Department of Energy, Office of Science, High-Energy Physics Program Office and Illinois Institute of Technology Department of Physics. 

IH led the development of the work presented in this paper. RL and RK provided supporting checks to calculations in this work and shaped the narrative of the manuscript. RL, RK, and IH drafted the manuscript. EFF and LH provided regular feedback over the course of the included work. All authors helped in preparation of this paper on the presented result.

\appendix

\section{Elastic Constants}

In Table~\ref{table:Elastics_Constants} we provide numerical values of the elastic constants used in this work. The second-order elastic constants are used to model caustic propagation, and are also used to calculate the second-order parameters \(\mu\) and \(\lambda\) needed for calculation of the anharmonic decay coefficient \(A\). The third-order elastic constants are used in finding \(\alpha\), \(\beta\), and \(\gamma\), also needed for calculating the anharmonic decay coefficient. All coefficients are given in (pressure) units of GPa. Sources for the elastic constants are found in the rightmost column. Standard third-order elastic constants for CaWO\(_{4}\) were not found.

\label{sec:AppendixA}
\begin{table*}[t]
\caption{Values of the second and third-order elastic constants in units of pressure (GPa) used in the calculation for anharmonic downconversion. We note that these are not all taken at the same temperature.}
\centering
\begin{tabular}{m{1.8cm}m{6.2cm}m{6.2cm}m{1.2cm}} \toprule \toprule
    {Material}  & \thead {Second-order \\ C\(_{11}\),C\(_{12}\),C\(_{44}\) [GPa]} & \thead{Third-order \\ C\(_{111}\),C\(_{112}\),C\(_{123}\),C\(_{144}\),C\(_{166}\),C\(_{456}\) [GPa]} & Refs.  \\ \midrule 
    Si & [167.8,65.2,80.1] & [-880,-515,27,74,-385,-40] & \cite{BerkeShort}  \\ 
    Ge & [126,44,67] & [-760,-410,-70,0,-310,-46] & \cite{Third_and_Second_Ge}   \\
    GaAs & [121.1,57.4,59.5] & [-620,-392,-62,8,-274.3,-43] &\cite{PhysRevB.32.5245_GaAs_Second_Order}  \\
    CaF\(_{2}\) & [174.5,56.4,35.93]  & [-1246,-400,-254,-124,-214,-75] & \cite{CaF2_Second_order_and_Third}\\
    LiF & [111.2,42.0,62.8] & [-1646,-252,92,98,-261,97] & \cite{LiF_Third_Order}  \\
\toprule    
    & \thead{Second-order \\ C\(_{11}\),C\(_{12}\),C\(_{13}\),C\(_{14}\),C\(_{34}\),C\(_{44}\) [GPa] }& \thead {Third-order \\ C\(_{111}\),C\(_{112}\),C\(_{113}\),C\(_{114}\),C\(_{123}\),C\(_{124}\),C\(_{133}\),C\(_{134}\),\\ C\(_{144}\),C\(_{155}\), C\(_{222}\),C\(_{333}\),C\(_{344}\),C\(_{444}\) [GPa]} &   \\
    Al\(_{2}\)O\(_{3}\) & [495, 161.1, 111.1, -22.2, 499.9,-147.7] & [-394.6,-112.7,-91.9,8.1,-20.4,-5.5,-96.4,-7.8,-38.2,-106.2,-453.4,-312.8,-113.1,4.1] & \cite{Second_Order_Al2O3,Third_order_Sapphire}   \\
 \toprule   
   & \thead{Second-order \\ C\(_{11}\),C\(_{12}\),C\(_{13}\),C\(_{16}\),C\(_{33}\),C\(_{44}\),C\(_{66}\) [GPa]} & \thead {Third-order \\ C\(_{111}\),C\(_{112}\),C\(_{113}\), C\(_{123}\),C\(_{133}\),C\(_{114}\),C\(_{155}\),C\(_{166}\),C\(_{333}\),\\ C\(_{344}\),C\(_{366}\), C\(_{456}\) [GPa]} & \\
    CaWO\(_{4}\)  & [151.5,65.6,45,18.8,134,35.4,40] & ------  & \cite{MsallCaWO4Experiment}  \\
   
\bottomrule \bottomrule
\end{tabular}
\label{table:Elastics_Constants}
\end{table*}

\section{Density of States}
\label{sec:AppendixB}

\begin{figure}[t!]
\includegraphics[width=\linewidth,height = 6.0cm]{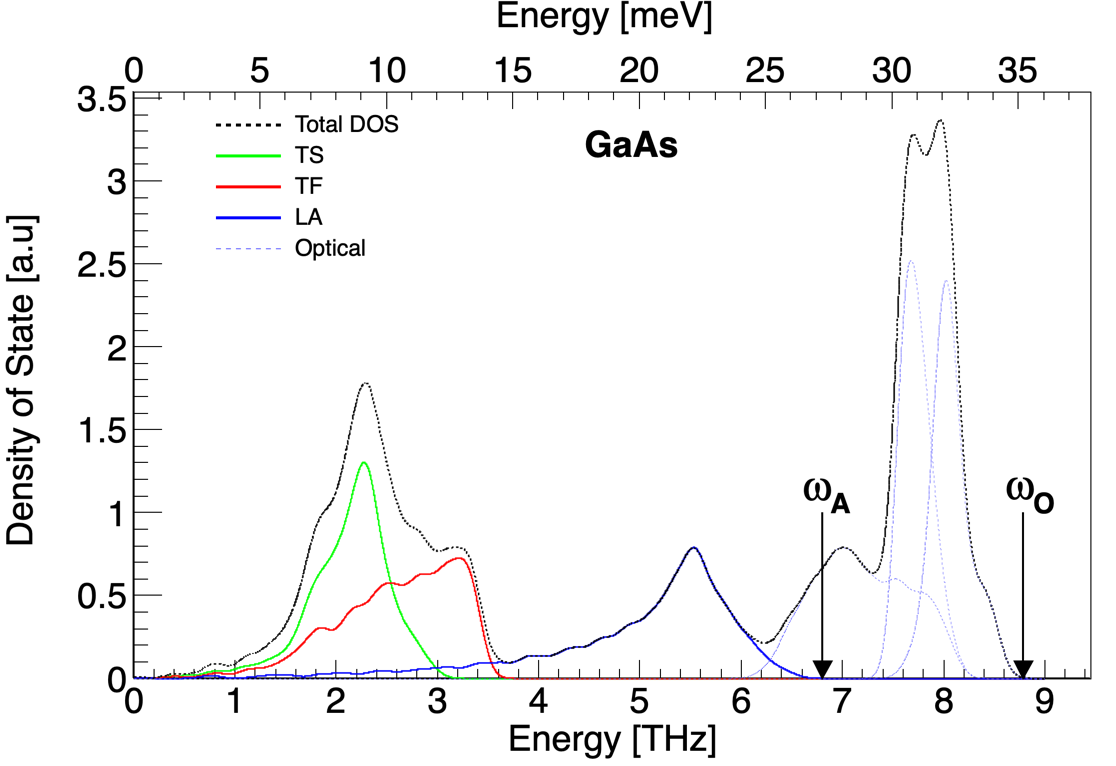}
\caption{Partial density of states of GaAs for transverse slow (green), transverse fast (red) and longitudinal (blue) acoustic phonons (solid lines). The blue dashed  curves to the right of the acoustic curves show the partial contribution of optical phonon channels to the DOS. The total density of states is the black line. $\omega_{A}$ corresponds to the maximum acoustic phonon energy and  $\omega_{O}$ is the maximum optical phonon energy.}
\label{fig:DOS_GaAs}
\end{figure}

\begin{figure}[t!]
\includegraphics[width=\linewidth,height = 6.0cm]{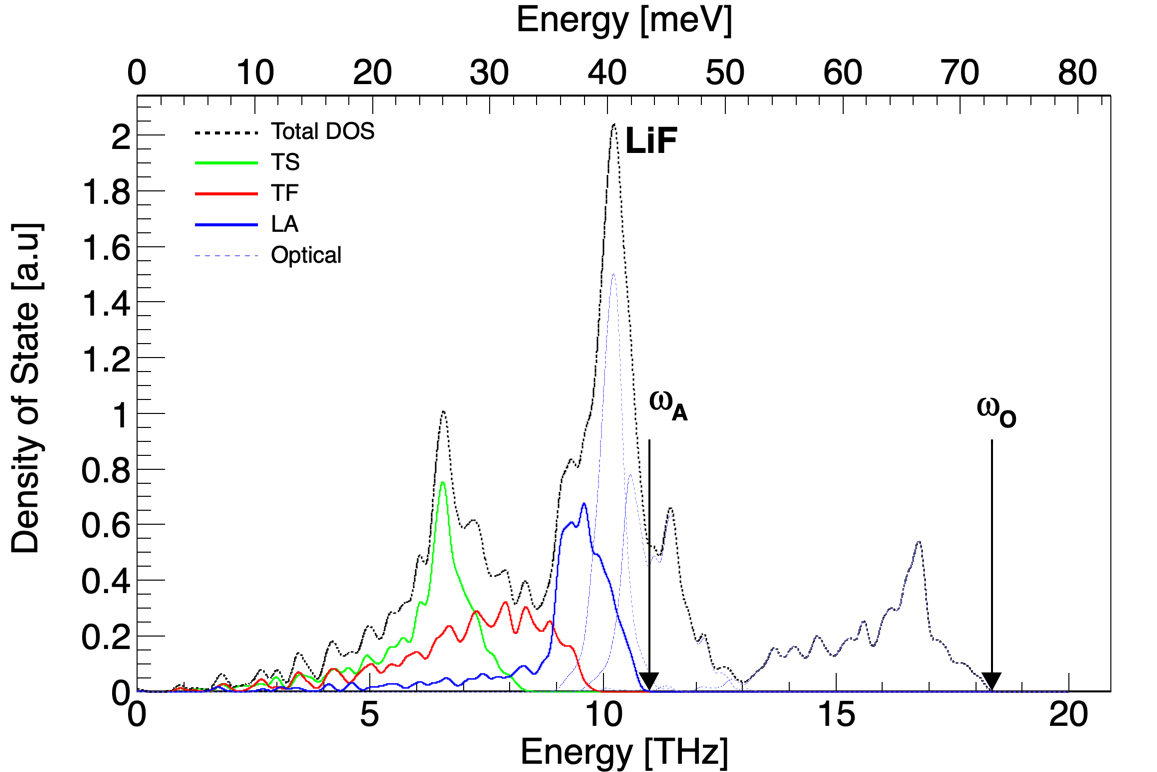}
\caption{Partial density of states of LiF for transverse slow (green), transverse fast (red) and longitudinal (blue) acoustic phonons (solid lines). The blue dashed curves to the right of the acoustic curves show the partial contribution of optical phonon channels to the DOS. The total density of states is the black line. $\omega_{A}$ corresponds to the maximum acoustic phonon energy and  $\omega_{O}$ is the maximum optical phonon energy.}
\label{fig:DOS_LiF}
\end{figure}

\begin{figure}[t!]
\includegraphics[width=\linewidth,height = 6.0cm]{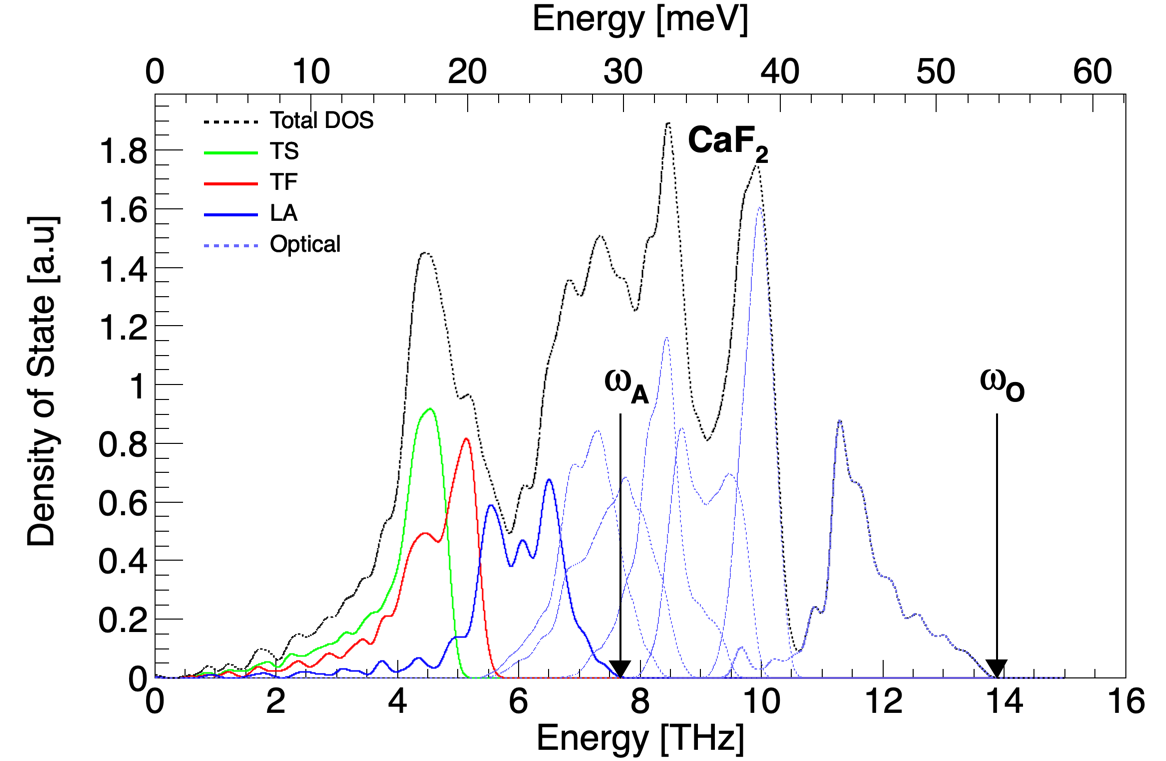}
\caption{Partial density of states of CaF\(_{2}\) for transverse slow (green), transverse fast (red) and longitudinal (blue) acoustic phonons (solid lines). The blue dashed curves to the right of the acoustic curves show the partial contribution of optical phonon channels to the DOS. The total density of states is the black line. $\omega_{A}$ corresponds to the maximum acoustic phonon energy and  $\omega_{O}$ is the maximum optical phonon energy.}
\label{fig:DOS_CaF2}
\end{figure}

We display phonon density of states curves for GaAs (Figure~\ref{fig:DOS_GaAs}), LiF (Figure~\ref{fig:DOS_LiF}), and CaF\(_{2}\) (Figure~\ref{fig:DOS_CaF2}). These curves have significantly reduced complexity in their optical branches relative to Al\(_{2}\)O\(_{3}\) due to the smaller number of atoms in a unit cell. The Debye Frequency \(\omega_{O}\) is used in G4CMP. 
The jitter in the DOS values at energies below approximately 5~THz is due to low-statistics in the numbers of points computed during the dispersion calculation. However, for LDOS, FTDOS, and STDOS values in Table~\ref{table:FractionalDOS}, a separate simulation with increased \(\vec{k}\)-space point density was used to avoid these fluctuations.

\begin{figure}[t!]
\includegraphics[width=\linewidth,height = 6.0cm]{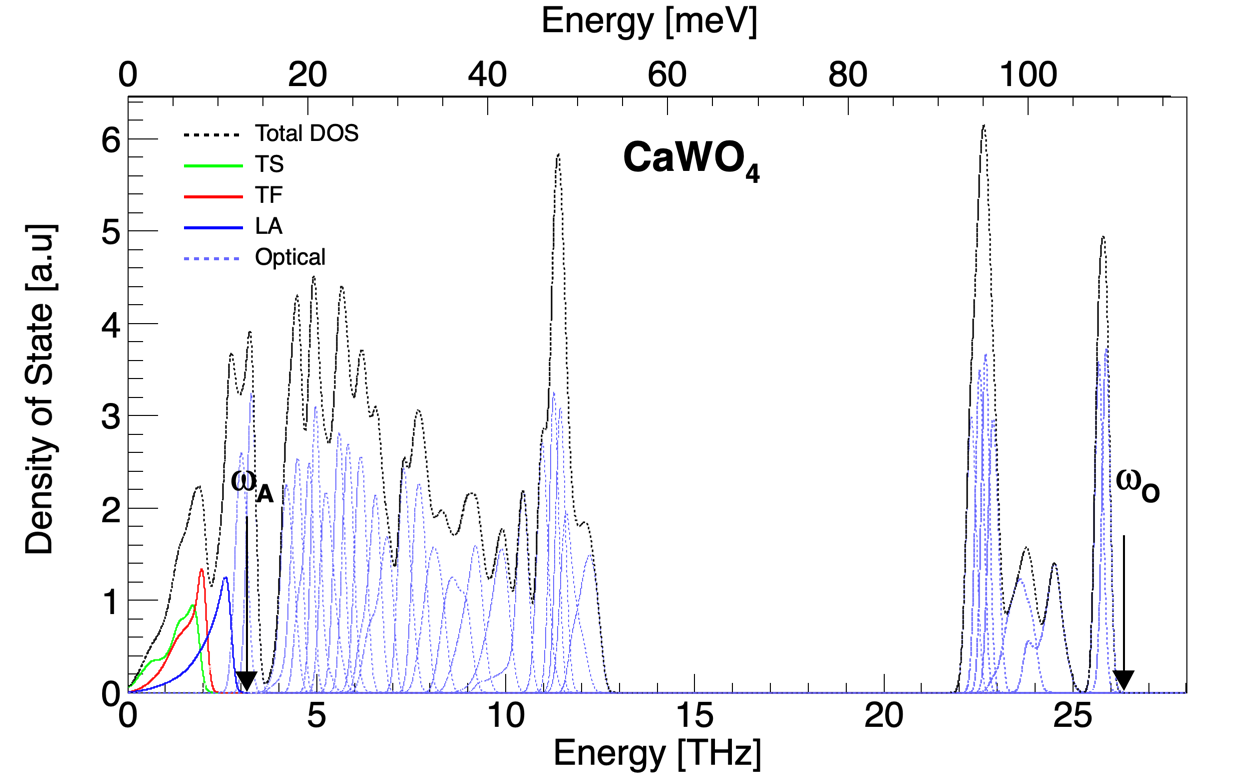}
\caption{Partial density of states of CaWO\(_{4}\) for transverse slow (green), transverse fast (red) and longitudinal (blue) acoustic phonons (solid lines). The blue dashed curves to the right of the acoustic curves show the partial contribution of optical phonon channels to the DOS. The total density of states is the black line. $\omega_{A}$ corresponds to the maximum acoustic phonon energy and  $\omega_{O}$ is the maximum optical phonon energy.}
\label{fig:DOS_CaWO4}
\end{figure}

\bibliography{main}{}

\end{document}